\documentclass[aps,preprint]{revtex4}
\usepackage{amsfonts}
\usepackage{amsmath}
\usepackage{epsfig}

\newtheorem{theorem}{Theorem}

\newtheorem{proposition}[theorem]{Proposition}

\input{tcilatex}

\begin{document}

\title{Determinant solution for the Totally Asymmetric Exclusion Process
with parallel update.}
\author{A.M. Povolotsky$^{1,2,}$}
\email{povolotsky@stp.dias.ie}
\author{V.B. Priezzhev$^{2,}$}
\email{priezzvb@thsun1.jinr.ru}
\affiliation{$^1$School of Theoretical Physics, Dublin Institute for Advanced Studies,
Dublin, Ireland,\\
$^{2}$Bogoliubov Laboratory of Theoretical Physics, Joint Institute for
Nuclear Research, 141980 Dubna, Russia}

\begin{abstract}
We consider the totally asymmetric exclusion process in discrete time with
the parallel update. Constructing an appropriate transformation of the
evolution operator, we reduce the problem to that solvable by the Bethe
ansatz. The non-stationary solution of the master equation for the infinite
1D lattice is obtained in a determinant form. Using a modified combinatorial
treatment of the Bethe ansatz, we give an alternative derivation of the
resulting determinant expression.
\end{abstract}

\maketitle


\address{$^1$Bogoliubov Laboratory of Theoretical physics, Joint Institute for Nuclear Research, Dubna, Russia,\email{priezzvb@thsun1.jinr.ru}
\\
$^2$School of Theoretical Physics, Dublin Institute for Advanced Studies, Dublin, Ireland }

\section{Introduction}

The first result on the Bethe ansatz solutions of the asymmetric simple
exclusion process (ASEP) is to be probably attributed to the short remark by
Dhar \cite{dhar}. The detailed calculations were later given in a seminal
paper of Gwa and Spohn \cite{gwa spohn}, where the issue about the scaling
of the next to the largest eigenvalues of the evolution operator was
addressed. The subject has attracted a significant attention since then and
a number of the exact results has become available. Among the results
obtained with the Bethe ansatz, there are the scaling function describing
the crossover of the next to the largest eigenvalue from the
Kardar-Parisi-Zhang to the Edwards-Wilkinson regime \cite{kim}, the large
deviation function for the distance travelled by a particle \cite{derrida
lebowitz},\cite{lee kim}. These results describe the behaviour of the system
in the large time limit. For the finite time evolution, we first of all
should mention the Sch\"{u}tz's determinant solution of the master equation
for the ASEP on the infinite lattice \cite{schutz}: the conditional
probability of an arbitrary particle configuration, given an initial
particle configuration, was obtained. This solution was later generalized to
the ring geometry \cite{priezzhev}, to the ASEP with particle dependent
velocities \cite{rakos schutz 1} and used to evaluate the distribution of
the current of particles across the arbitrary bond \cite{rakos schutz} at
the infinite lattice.

All the above mentioned results concern the continuous time ASEP, where the
jump of a particle is described by the Poisson process. The continuous time
solution can be straightforwardly extended to the discrete time dynamics
with random sequential update \cite{lee kim}. The Bethe ansatz solution for
the ASEP with backward ordered update \cite{open case 1} was given in \cite%
{brankov priezzhev shelest}. The history of studies of the ASEP with fully
parallel update rises probably from the 90s, when it was widely discussed as
a simplest version of the Nagel-Schreckenberg traffic model with maximal
velocity of cars: $v_{\max }=1$ \cite{nagel schreckenberg}. In that case the
approximate mean-field method proposed to find the stationary measure of the
traffic flow models \cite{schreckenberg schadschneider nagel ito} resulted
in the expression, which becomes exact in the thermodynamic limit. Later it
was realized that the mapping to the zero range process (ZRP) allows one to
represent the stationary measure of the parallel ASEP at the ring in form of
the product measure \cite{evans}. In the case of open boundary conditions
the stationary measure was obtained in form of the matrix product \cite{open
case 1},\cite{open case 2},\cite{open case 3}. At the same time, the
description of the time evolution of the ASEP with parallel update remained
challenging unsolved problem for a long time. Recently, it has been shown
that the parallel ASEP is a particular case of the general stochastic
process with parallel update, which is solvable by the Bethe ansatz \cite%
{povolotsky mendes}.

In present article we use the Bethe ansatz to obtain the solution of the
master equation for the ASEP with parallel update on an infinite lattice.
Given an initial configuration, we obtain the probability of an arbitrary
configuration of particles. The solution is expressed in terms of the
determinant of the matrix, which depends on the initial and final
configurations of particles and time. Though the general idea of the
solution is similar to that of \cite{schutz}, the parallel dynamics leads to
peculiarities, which make a generalization of previous results nontrivial.
Below we give two independent derivations of the final result using two
different treatments of the Bethe ansatz: the analytic representation based
on the expansion of the solution over the Bethe ansatz eigenbasis and the
geometric one based on the direct enumeration of particle trajectories.
While the former reveals the peculiarities of the eigenspace of the
evolution operator, the latter clarifies combinatorial aspects of the
problem. Of course both methods lead to the same final result.

The article is organized as follows. In the Section II, we formulate the
model and sketch the results of the article. In the Sections III and IV we
derive the solution using the analytic and geometric Bethe ansatz
respectively. In the Section V we summarize the results and discuss
perspectives.

\section{Formulation of the model and results.}

Let us consider $P$ particles on 1D infinite lattice. Below we imply that
the particles move from left to right, which will be also referred to as a
forward direction. At each step of the discrete time each particle may take
one step forward with probability $v$ or stay with probability $(1-v)$
provided that the next site is vacant. If the next site is occupied, the
particle stays with probability $1$. All sites are updated simultaneously at
each time step. We define a configuration $X$ of particles by a set of their
coordinates $X=\left\{ x_{1},x_{2},\ldots ,x_{P}\right\} $. Below we imply
them to be ordered: 
\begin{equation}
-\infty <x_{1}<x_{2}<\cdots <x_{P}<\infty  \label{x_1<x_2<...<x_M.}
\end{equation}%
The probability $P_{t}(X)$ for the system to be in a configuration $X$ at
time $t$ obeys the Markov equation%
\begin{equation}
P_{t+1}(X)=\sum_{\left\{ X^{\prime }\right\} }T(X,X^{\prime
})P_{t}(X^{\prime }),
\end{equation}%
where $T(X,X^{\prime })$ is the probability of the transition from $%
X^{\prime }$ to $X$ for one time step. The transition probabilities $%
T(X,X^{\prime })$ are defined by the above dynamical rules.

The goal of the article is to show that the conditional probability, $%
P(X,t|X^{0},0)$, for the system to be in a configuration $X=\left\{
x_{1},x_{2},\ldots ,x_{P}\right\} $ at time $t$, given it was in a
configuration $X^{0}=\left\{ x_{1}^{0},x_{2}^{0},\ldots ,x_{P}^{0}\right\} $
at time $0$, is the ratio of two determinants 
\begin{equation}
P(X,t|X^{0},0)=\frac{\det \mathbf{F}\left( X,X^{0},t\right) }{\det \mathbf{F}%
\left( X,X,0\right) }\mathbf{.}  \label{P(X,t|X_0,0)}
\end{equation}%
The matrix elements of the matrix $\mathbf{F}\left( X,Y,t\right) $, which
depend on two configurations $X=\left\{ x_{1},x_{2},\ldots ,x_{P}\right\} $
and $Y=\left\{ y_{1},y_{2},\ldots ,y_{P}\right\} $ and on time $t$, are%
\begin{equation}
\left( \mathbf{F}\left( X,Y,t\right) \right) _{i,j}=f(i-j,x_{i}-y_{j},t),
\label{F(x,y,t)_i.j}
\end{equation}%
where the function $f\left( a,b,t\right) $ is expressed in terms of the
Gauss hypergeometric functions:

\begin{equation}
f(a,b,t)=\left( 1-v\right) ^{t}\left( -1\right)^{a} 
\begin{cases}
\begin{array}{ll}
\left( \frac{v}{v-1}\right) ^{b}\frac{(-t-a)_{b}}{b!}\left. _{2}F_{1}\right.
\left( 
\begin{array}{c}
a,-t-a+b \\ 
b+1%
\end{array}%
;\frac{v}{v-1}\right) & b>0 \\ 
\frac{\left( a\right) _{-b}}{\left( -b\right) !}\left. _{2}F_{1}\right.
\left( 
\begin{array}{c}
a-b,-t-a \\ 
-b+1%
\end{array}%
;\frac{v}{v-1}\right) & b\leq 0%
\end{array}
.%
\end{cases}
\label{f(a,b,t)}
\end{equation}%
The notation $\left( a\right) _{n}$ means the shifted factorial $\left(
a\right) _{n}=a(a+1)\cdots \left( a+n-1\right) $.

\section{\protect\bigskip Analytic approach.}

\subsection{Hilbert space and the vector of state.}

Let us introduce the Hilbert space supplied with the complete left and right
bases consisting of vectors $\left\langle X\right\vert $ and $\left\vert
X\right\rangle $ respectively, where $X$ runs over all particle
configurations, with inner product 
\begin{equation}
\left\langle X|X^{\prime }\right\rangle =\delta (X,X^{\prime }).
\end{equation}%
The state of the system at any time step can be associated with the state
vector%
\begin{equation}
\left\vert P_{t}\right\rangle =\sum_{\left\{ X^{\prime }\right\}
}P_{t}(X)\left\vert X\right\rangle .
\end{equation}%
In terms of the state vectors, the master equation takes a simple operator
form 
\begin{equation}
\left\vert P_{t+1}\right\rangle =\mathbf{T}\left\vert P_{t}\right\rangle ,
\end{equation}%
where the evolution operator $\mathbf{T}$ is defined as follows%
\begin{equation}
\mathbf{T=}\sum_{\{X\},\{X^{\prime }\}}\left\vert X\right\rangle
T(X,X^{\prime })\left\langle X^{\prime }\right\vert .
\end{equation}%
The conditional probability $P(X,t|X^{0},0)$ can be represented as the
matrix element 
\begin{equation}
P(X,t|X^{0},0)=\left\langle X\right\vert \mathbf{T}^{t}\left\vert
X^{0}\right\rangle .  \label{T^t}
\end{equation}%
To evaluate matrix elements, we construct the set of left eigenvectors $%
\left\vert B_{Z}\right\rangle $ of the operator $\mathbf{T}$%
\begin{equation}
\mathbf{T}\left\vert B_{Z}\right\rangle =\Lambda (Z)\left\vert
B_{Z}\right\rangle  \label{T|Z>=Lambda(Z)|Z>}
\end{equation}%
and the adjoint set of right eigenvectors $\left\langle \overline{B}%
_{Z}\right\vert $ 
\begin{equation}
\left\langle \overline{B}_{Z}\right\vert \mathbf{T}=\Lambda (Z)\left\langle 
\overline{B}_{Z}\right\vert ,  \label{<z|T=lambda<z|}
\end{equation}%
which, as will be seen below, are parametrized by the $P$-fold parameter $Z$%
. We will prove that the set forms a complete basis, i.e. provides the
expansion of the identity operator 
\begin{equation}
\sum\limits_{Z}\left\langle X|B_{Z}\right\rangle \left\langle \overline{B}%
_{Z}|X^{\prime }\right\rangle =\left\langle X|X^{\prime }\right\rangle .
\label{<x|x'>}
\end{equation}%
As a result, the transition probability (\ref{T^t}) we are looking for can
be reduced to the evaluation of the sum 
\begin{equation}
\left\langle X\right\vert \mathbf{T}^{t}\left\vert X^{0}\right\rangle
=\sum\limits_{Z}\left\langle X|\mathbf{T}^{t}|B_{Z}\right\rangle
\left\langle \overline{B}_{Z}|X^{0}\right\rangle =\sum\limits_{Z}\Lambda
^{t}\left( Z\right) \left\langle X|B_{Z}\right\rangle \left\langle \overline{%
B}_{Z}|X^{0}\right\rangle .  \label{<x|T^t|x_0>}
\end{equation}

\subsection{Bethe ansatz.}

The solution of the eigenproblem (\ref{T|Z>=Lambda(Z)|Z>}) corresponding to
the master equation of the general integrable ASEP-ZRP model with parallel
update was given in \cite{povolotsky mendes}. Here we sketch the solution
for the particular case of the parallel ASEP defined above.

For further purposes it is convenient to introduce a notion of cluster of
particles. A sequence of $n$ occupied sites situated one after another with
two empty sites on the ends without any other empty sites between them will
be referred to as a cluster of the length $n$. A single isolated particle is
the cluster of the unit length. During the evolution an isolated particle
can approach and join a cluster from behind resulting in decrease of the
number of clusters at the lattice by one. The opposite process is when a
cluster splits up by a first particle detaching the cluster ahead creating
an isolated particle and increasing the number of clusters by one. If a
particle jumps between two clusters, which are separated by only one empty
site, the number of clusters does not change.

\begin{quotation}
\textit{The explicit form of the} \textit{transition probability }$%
T(X,X^{\prime })$\textit{\ is a product of factors, each corresponding to a
particular cluster of particles in the initial configuration }$X^{\prime }$%
\textit{. The value of these factors is either }$v$\textit{\ or }$(1-v)$%
\textit{\ depending on whether or not the first particle of a given cluster
jumps during the transition from }$X^{\prime }$\textit{\ to }$X.$\textit{\ } 
\textit{\ }
\end{quotation}

\begin{equation}
T(X,X^{\prime })=\prod\limits_{i=1}^{\mathcal{N}_{c}\left( X^{\prime
}\right) }\left( 1-v\right) ^{1-m_{i}}v^{m_{i}}
\end{equation}%
Here $m_{i}=0,1$ is the number of particles hopping from $i$-th cluster of $%
X^{\prime }$.

To proceed with the solution, we introduce a transformation \cite{povolotsky
mendes} given by the diagonal in the configurational basis operator $%
\mathcal{D}$ 
\begin{equation*}
\mathcal{D}=\sum\limits_{\{X\}}\frac{\left\vert X\right\rangle \left\langle
X\right\vert }{W(X)}.
\end{equation*}%
where the wight $W(X)$ is the stationary measure of the process. For the
fixed number of particles at the infinite lattice, it can be given in terms
of the number of clusters $\mathcal{N}_{c}\left( X\right) $ in configuration 
$X$ 
\begin{equation}
W\left( X\right) =\left( 1-v\right) ^{P-\mathcal{N}_{c}\left( X\right) }.
\label{W(X)}
\end{equation}%
Technically, it is more convenient to solve first the right eigenproblem 
\begin{equation}
\mathbf{T}_{0}\left\vert B_{Z}^{0}\right\rangle =\Lambda (Z)\left\vert
B_{Z}^{0}\right\rangle ,  \label{Lambda P^0=TP^0}
\end{equation}%
for the operator $\mathbf{T}_{0}$ defined as follows 
\begin{equation*}
\mathbf{T}_{0}=\mathcal{D}\mathbf{T}\mathcal{D}^{-1}
\end{equation*}%
rather than for the operator $\mathbf{T}$, which have the same eigenvalue
but different eigenvectors. Then the eigenvectors $\left\vert
B_{Z}\right\rangle $ of $\mathbf{T}$ can be related to eigenvectors $%
\left\vert B_{Z}^{0}\right\rangle $ of $\mathbf{T}_{0}$ 
\begin{equation}
\left\vert B_{Z}\right\rangle =\mathcal{D}^{-1}\left\vert
B_{Z}^{0}\right\rangle .  \label{<x|B_z>}
\end{equation}%
Explicitly the matrix elements 
\begin{equation}
T_{0}(X,X^{\prime })=\frac{W(X^{\prime })}{W(X)}T(X,X^{\prime }).
\label{T=T W(x')/W(x)}
\end{equation}%
can be constructed in terms of the cluster structure of final configuration $%
X$ according to the following rule:

\begin{quotation}
\textit{If a cluster of particles in }$X$\textit{\ receives a particle from
behind, it contributes the factor }$v$\textit{\ to }$T_{0}(X,X^{\prime })$%
\textit{\ and }$(1-v)$\textit{\ otherwise. }%
\begin{equation}
T_{0}(X,X^{\prime })=\prod\limits_{i=1}^{\mathcal{N}_{c}\left( X\right)
}\left( 1-v\right) ^{1-k_{i}}v^{k_{i}}  \label{T0}
\end{equation}
\end{quotation}

\noindent where $k_{i}=0,1$ is the number of particles approaching $i$-th
cluster of $X$ from behind. Indeed, the event when an isolated particle
joins a cluster, the first particle of which stays, contributes $v(1-v)~$\
to $T(X,X^{\prime })$. At the same time, as the two clusters merge into one,
it follows from (\ref{W(X)},\ref{T=T W(x')/W(x)}) that the additional factor
of $\left( 1-v\right) ^{-1}$ should be incorporated into $T_{0}(X,X^{\prime
})$, so that the final contribution will be $v$ (see Fig \ref{transition
probabilities}). 
\begin{figure}[tbp]
\unitlength=1mm \makebox(110,45)[cc] {\psfig{file=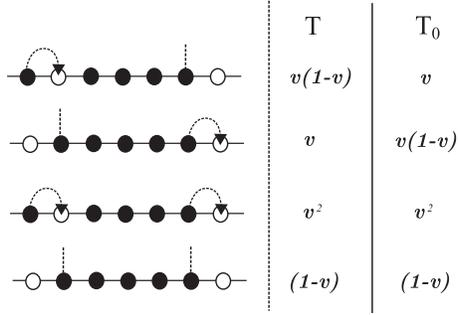,width=60mm}}
\caption{The transformation of the transition probabilities.}
\label{transition probabilities}
\end{figure}
If the first particle of a cluster of the initial configuration jumps
forward, while no particle is received from behind, the probability of the
jump $v$ enters into $T(X,X^{\prime })$. However, as one cluster in the
initial configuration has split up into two clusters in the final one, the
factor $\left( 1-v\right) $ should be added in $T_{0}(X,X^{\prime })$. The
processes, where the both first and the last particles of the cluster stay,
proceed without change of the number of clusters $\mathcal{N}_{c}\left(
X\right) $, and hence their contributions to $T(X,X^{\prime })$ and to $%
T_{0}(X,X^{\prime })$ are equal as well as when the cluster of the initial
configuration receives a particle from behind while its first particle
detaches it at the front. The expression (\ref{T0}) can also be obtained in
more formal way using the correspondence between the ASEP to the ZRP as it
was done in \cite{povolotsky mendes}. As a result, (\ref{Lambda P^0=TP^0})
can be written in the form%
\begin{eqnarray}
\Lambda \left( Z\right) \langle \ldots ,\underset{i-th~cluster}{\underbrace{%
x,x+1,\ldots ,x+n_{i}}},\ldots |B_{Z}^{0}\rangle &=&\sum\limits_{\left\{
k_{i}\right\} }\prod\limits_{i=1}^{\mathcal{N}_{c}\left( X\right)
}v^{k_{i}}\left( 1-v\right) ^{1-k_{i}}  \notag \\
&&\langle \ldots ,x-k_{i},x+1,\ldots ,x+n_{i},\ldots |B_{Z}^{0}\rangle ,
\label{masterinteract}
\end{eqnarray}%
where the index $i$ in the product runs over all the clusters of particles
of the configuration $X$. The number $k_{i}$ takes on the value of $1$ or $0$
depending on whether or not the $i$-th cluster receives the particle from
behind. The summation is performed over all possible sets $\left\{
k_{i}\right\} $. The $i$-th cluster consisting of $n_{i}$ particles is
written explicitly in the argument\ in formula (\ref{masterinteract}).

As usual, the strategy of the Bethe ansatz solution is the following. We
note that when all the particles are at least one empty site apart from each
other, they behave like noninteracting, i.e. each particle decides whether
it is going to jump independently of the location of the others. In other
words, if the coordinates of particles $\left\{ x_{1},\ldots ,x_{P}\right\} $
of the configuration $X$ satisfy the inequality $\left( x_{i}+1\right)
<x_{i+1}$ for every $i$, $|B_{Z}^{0}\rangle $ obeys the following equation 
\begin{eqnarray}
\Lambda \left( Z\right) \left\langle x_{1},\ldots
,x_{P}|B_{Z}^{0}\right\rangle &=&\sum_{\left\{ k_{i}\right\}
}\prod_{i=1}^{P}v^{k_{i}}\left( 1-v\right) ^{1-k_{i}}  \notag \\
&&\times \left\langle x_{1}-k_{1},\ldots ,x_{P}-k_{P}|B_{Z}^{0}\right\rangle
.  \label{masterfree}
\end{eqnarray}%
Here the product is taken over all $P$ particles, and the summation is over
all $k_{i}$, which take on the values of $1$ and $0$ depending on whether or
not the $i$-th particle decided to jump. The necessary trick is to find a
way for this equation to be satisfied in the whole set of particle
configurations including the domains where the equality $\left(
x_{i}+1\right) =x_{i+1}$ takes place. If we consider formally the equation (%
\ref{masterfree}) for such a configuration, the terms like $\left\langle
\ldots ,x,x,\ldots |B_{Z}^{0}\right\rangle $ will appear. They are beyond
the allowed domain\ (\ref{x_1<x_2<...<x_M.}) where the sets of particle
coordinates are defined. Therefore they should be redefined and cancelled so
that the difference with (\ref{masterinteract}) disappears. Particularly,
one needs to make sure that the result of the one cluster summation in (\ref%
{masterinteract}) coincides with the corresponding term of the free equation,%
\begin{eqnarray}
&&\!\!\!\!\!\!\!\!\!\!v\left\langle \ldots ,x-1,x+1,\ldots ,x+n,\ldots
|B_{Z}^{0}\right\rangle +\left( 1-v\right) \left\langle \ldots ,x,x+1,\ldots
,x+n,\ldots |B_{Z}^{0}\right\rangle  \notag \\
&=&\sum_{\left\{ k_{i}\right\} }\prod_{i=0}^{n}v^{k_{i}}\left( 1-v\right)
^{1-k_{i}}\left\langle \ldots ,x-k_{0},x+1-k_{2},\ldots ,x+n-k_{n},\ldots
|B_{Z}^{0}\right\rangle .
\end{eqnarray}%
This relation can be satisfied with the only constraint on $B_{Z}^{0}\left(
x_{1},\ldots ,x_{P}\right) $%
\begin{align}
v& \left[ \left\langle \ldots ,x-1,x,\ldots |B_{Z}^{0}\right\rangle
-\left\langle \ldots ,x-1,x+1\ldots |B_{Z}^{0}\right\rangle \right]  \notag
\\
& +(1-v)\left[ \left\langle \ldots ,x,x,\ldots |B_{Z}^{0}\right\rangle
-\left\langle \ldots ,x,x+1,\ldots |B_{Z}^{0}\right\rangle \right] =0,
\label{constraint}
\end{align}%
which can be checked by a direct calculation for the size of the cluster $%
n=2 $, and then proven by induction for larger $n$ (see \cite{povolotsky
mendes}).

After that one can look for $B_{Z}^{0}\left( x_{1},\ldots ,x_{P}\right) $ in
the form of the Bethe ansatz,%
\begin{equation}
\left\langle x_{1},\ldots ,x_{P}|B_{Z}^{0}\right\rangle =\sum_{\left\{
\sigma \right\} }A_{\sigma _{1}\ldots \sigma _{P}}z_{\sigma
_{1}}^{-x_{1}}\ldots z_{\sigma _{P}}^{-x_{P}},  \label{Bethe ansatz}
\end{equation}%
parametrized by the parameter $Z,$ which is a set of $P$\ complex numbers $%
Z=\left\{ z_{1},\ldots ,z_{P}\right\} $. Here the summation is performed
over all permutations $\sigma $ of the numbers $1,\ldots ,P$. \ The
substitution of (\ref{Bethe ansatz}) to the free equation (\ref{masterfree})
yields the expression for the eigenvalues in terms of parameters $\left\{
z_{1},\ldots ,z_{P}\right\} $%
\begin{equation}
\Lambda \left( Z\right) =\left( 1-v\right) ^{P}\prod\limits_{i=1}^{P}\left(
1+\lambda z_{i}\right) ,  \label{Lambda(Z)}
\end{equation}%
where%
\begin{equation}
\lambda =\frac{v}{1-v}.  \label{lambda}
\end{equation}%
The substitution of the Bethe ansatz into the constraint (\ref{constraint})
results in the relation between the amplitudes $A_{\sigma _{1}\ldots \sigma
_{P}}$, which differ from each other only by two neighboring indices permuted%
\begin{equation}
\frac{A_{\ldots ij\ldots }}{A_{\ldots ji\ldots }}=-S\left(
z_{i},z_{j}\right) ,  \label{A_ji=A_ij...}
\end{equation}%
where 
\begin{equation}
S\left( z_{i},z_{j}\right) \equiv \frac{1-1/z_{i}}{1-1/z_{j}}\frac{1+\lambda
z_{j}}{1+\lambda z_{i}}.  \label{s-matrix}
\end{equation}

To obtain an explicit form of the amplitudes $A_{\sigma }$, consider a
particular permutation $\left( \sigma _{1},\ldots ,\sigma _{P}\right) $. Let
us associate the factor $\xi _{i}\xi _{j}^{-1}$ to an elementary
transposition $\left( \ldots i,j\ldots \right) \rightarrow \left( \ldots
j,i\ldots \right) $. Apparently the power of $\xi _{i}$ will then coincide
with the position of $i$ in the permutation. Thus, we have the following
correspondence%
\begin{equation}
\left( 
\begin{array}{c}
1,\ldots ,P \\ 
\sigma _{1},\ldots ,\sigma _{P}%
\end{array}%
\right) .\rightarrow \xi _{\sigma _{_{1}}}^{1-\sigma _{_{1}}}\xi _{\sigma
_{_{2}}}^{2-\sigma _{_{2}}}\ldots \xi _{\sigma _{P}}^{P-\sigma _{P}}.
\end{equation}%
From (\ref{A_ji=A_ij...}),we have%
\begin{equation}
\xi _{i}=-\frac{1+\lambda z_{i}}{1-1/z_{i}}.
\end{equation}%
and%
\begin{equation}
A_{\sigma _{1}\ldots \sigma _{P}}=\left( -1\right) ^{\mathcal{P}\left(
\left\{ \sigma \right\} \right) }\prod\limits_{i=1}^{P}\left( -\frac{%
1+\lambda z_{\sigma _{i}}}{1-1/z_{\sigma _{i}}}\right) ^{i-\sigma _{i}},
\label{A_sigma}
\end{equation}%
where $\mathcal{P}\left( \left\{ \sigma \right\} \right) $ is the parity of
a permutation $\left\{ \sigma \right\} $. The formulas (\ref{<x|B_z>},\ref%
{Bethe ansatz},\ref{A_sigma}) together with the definition of the weight (%
\ref{W(X)}) is all what we need to write the right eigenvector $\left\vert
B_{Z}\right\rangle $ as a set of its coordinates $\left\langle
X|B_{Z}\right\rangle $ in the configurational basis $\left\{ \left\vert
X\right\rangle \right\} $.

To construct the adjoint set of the left eigenvectors $\left\langle 
\overline{B}_{Z}\right\vert $ one has to solve (\ref{<z|T=lambda<z|}) or
equivalently the right eigenproblem for the transposed matrix $\mathbf{T}%
^{\ast }$%
\begin{equation}
\mathbf{T}^{\ast }\left\vert \overline{B}_{Z}\right\rangle =\Lambda
(Z)\left\vert \overline{B}_{Z}\right\rangle .  \label{T^T|Z>}
\end{equation}%
To this end, we note that in the original process corresponding to the
matrix $\mathbf{T}$, the first particle of a cluster contributes the factor $%
v$ to the transition probability $T(X,X^{\prime })$ if it jumps forward and $%
\left( 1-v\right) $ if it stays. In the adjoint process, which corresponds
to the transposed matrix $\mathbf{T}^{\ast }$ the particles move in the
opposite direction. The last particle of a cluster contributes the factor $v$
or $\left( 1-v\right) $ to a matrix element if it jumps approaching the
cluster or stays in it respectively. One can see that the action of the
operator $\mathbf{T}^{\ast }$ is similar, up to the inversion of the
coordinates, to the one of the operator $\mathbf{T}_{0}$, (\ref{T=T
W(x')/W(x)}). In other words, the matrices $\mathbf{T}_{0}$ and $\mathbf{T}%
^{\ast }$ are related by the identity 
\begin{equation}
\mathbf{T^{\ast }}=\mathbf{IT}_{0}\mathbf{I}  \label{detailed balance}
\end{equation}%
where $\mathbf{I}$ is the operator of the inversion of coordinates, $\mathbf{%
I}\left\vert x_{1,}\ldots ,x_{p}\right\rangle =\left\vert -x_{p},\ldots
,-x_{1}\right\rangle $. It is then straightforward to show that the solution
of the equation (\ref{T^T|Z>}), corresponding to the eigenvalue (\ref%
{Lambda(Z)}) is given by the Bethe ansatz 
\begin{equation}
\left\langle \overline{B}_{Z}|X\right\rangle =\frac{1}{P!}\sum_{\left\{
\sigma \right\} }\overline{A}_{\sigma _{1}\ldots \sigma _{P}}z_{\sigma
_{1}}^{x_{1}}\ldots z_{\sigma _{P}}^{x_{P}}  \label{<z|x>}
\end{equation}%
with the amplitudes $\overline{A}_{\sigma }$ obeying the relation%
\begin{equation}
\frac{\overline{A}_{\ldots ij\ldots }}{\overline{A}_{\ldots ji\ldots }}%
=-S\left( z_{j},z_{i}\right) .  \label{A tilde ij/A tilde ji}
\end{equation}%
where the $S$-matrix is the same as in the relation for $A_{\sigma }$, (\ref%
{s-matrix}), with the permuted arguments. As a result we obtain \ 
\begin{equation*}
A_{\sigma _{1}\ldots \sigma _{P}}=1/\overline{A}_{\sigma _{1}\ldots \sigma
_{P}}
\end{equation*}%
The constant factor $\left( P!\right) ^{-1}$ is introduced for further
convenience. Finally $\left\langle \overline{B}_{Z}|X\right\rangle $ and $%
\left\langle X|B_{Z}\right\rangle $ can be represented in the determinant
form%
\begin{eqnarray}
\left\langle X|B_{Z}\right\rangle &=&W(X)\det \mathbf{B,} \\
\left\langle \overline{B}_{Z}|X\right\rangle &=&\frac{1}{P!}\det \overline{%
\mathbf{B}}
\end{eqnarray}%
where the matrix elements $B_{ij}$ and $\overline{B}_{ij}$ of the matrices $%
\mathbf{B}$ and $\overline{\mathbf{B}}$ respectively are given by%
\begin{equation}
B_{ij}=1/\overline{B}_{ij}=\left( -\frac{1+\lambda z_{j}}{1-1/z_{j}}\right)
^{i-j}z_{j}^{-x_{i}}.  \label{B_ij}
\end{equation}

\subsection{Completeness of the eigenbasis and the transition probabilities.}

To proceed, we should define the range of the summation over $Z$ in (\ref%
{<x|x'>},\ref{<x|T^t|x_0>}). For the infinite lattice, the spectrum of $Z$
can be taken continuous, and the summation over $Z$ should be replaced by
the $P-$ fold contour integral 
\begin{equation}
\sum\limits_{Z}\rightarrow \prod\limits_{i=1}^{P}\oint \frac{dz_{i}}{2\pi 
\mathrm{i}z_{i}}.
\end{equation}%
The contour of the integration over each $z_{i}$ is closed around the points 
$z=0~$and $\ z=1$, so that the point $z=-1/\lambda $ stays outside, Fig.(\ref%
{contour}). 
\begin{figure}[tbp]
\unitlength=1mm \makebox(110,45)[cc] {\psfig{file=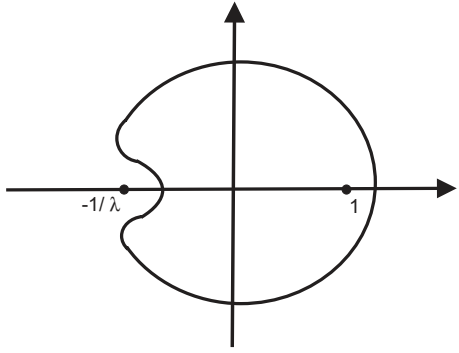,width=60mm}}
\caption{{}Contour of integration over the parameters $z_{1},\ldots ,z_{M}$.}
\label{contour}
\end{figure}
This is a generalization of the solution for the continuous time ASEP. In
the limit $\lambda \rightarrow 0$ the point $z=-1/\lambda $ goes to
infinity, and the situation becomes similar to the one studied in \cite%
{schutz}. Below, we prove the completeness of the eigenbasis, which follows
from (\ref{<x|x'>}). As soon as it is established, one can directly
substitute (\ref{Bethe ansatz},\ref{Lambda(Z)},\ref{A_sigma}) into the
expression for the probability (\ref{<x|T^t|x_0>}) and obtain 
\begin{equation}
\left\langle X\right\vert \mathbf{T}^{t}\left\vert X^{0}\right\rangle
=W(X)\sum_{\left\{ \sigma \right\} }\left( -1\right) ^{\sigma
}\prod\limits_{i=1}^{P}f(i-\sigma _{i},x_{i}-x_{\sigma _{i}}^{0},t)
\label{<x|T^t|x_0>-explicit form}
\end{equation}%
where $(-1)^{\sigma }$ is the sign of the permutation $\sigma $ and the
function $f(a,b,t)$ is defined as follows 
\begin{equation}
f(a,b,t)=\left( 1-v\right) ^{t}\oint \frac{dz}{2\pi \mathrm{i}z}\left(
1+\lambda z\right) ^{t}\left( -\frac{1+\lambda z}{1-1/z}\right) ^{a}z^{-b}.
\label{f(a,b,t=INT)}
\end{equation}%
We should note that initially there should be a double sum over permutations 
$\sigma $ and $\sigma ^{\prime }$ as the terms $\left\langle
X|B_{Z}\right\rangle $ and $\left\langle \overline{B}_{Z}|X\right\rangle $
contains one sum each. However, there is no any dependence on $\sigma
^{\prime }$ in the product of the factors $f(\sigma _{i}^{\prime }-\sigma
_{\sigma _{i}^{\prime }},x_{\sigma _{i}^{\prime }}-x_{\sigma _{\sigma
_{i}^{\prime }}}^{0},t)$ over all $i$-s. As a result we have $P!$ equal
terms under the sum over $\sigma ^{\prime }$, which just eliminate the
prefactor $1/\left( P!\right) $ in (\ref{<z|x>}) and the only one sum (\ref%
{<x|T^t|x_0>-explicit form}) over permutations remains.

Consider two configurations 
\begin{equation}
X=\left\{ x_{1},<x_{2},<\ldots ,<x_{P}\right\} ,Y=\left\{
y_{1},<y_{2},<\ldots ,<y_{P}\right\} .  \label{domain}
\end{equation}%
The explicit form of (\ref{<x|x'>}) is given by the particular case of
formulas (\ref{<x|T^t|x_0>-explicit form},\ref{f(a,b,t=INT)}) taken at $t=0$%
, $X^{0}=Y$. Consider a term of the sum corresponding to a particular
permutation $\sigma $, which is the product of $P$ integrals of the form.%
\begin{equation}
f(i-\sigma _{i},x_{i}-y_{\sigma _{i}},0)=\oint \frac{dz}{2\pi \mathrm{i}z}%
\left( -\frac{1+\lambda z}{1-1/z}\right) ^{i-\sigma _{i}}z^{-x_{i}+y_{\sigma
_{i}}}.  \label{integral1}
\end{equation}%
Apparently, the summand to be nonzero, the integrals for all $i$-s should be
nonzero simultaneously. The following statement can be proved.

\begin{proposition}
\label{prop} \textit{The expression (\ref{integral1}) yields a nonzero
result for all }$i=1,\ldots ,P$ \textit{only when }$X=Y\,$\textit{and only
for those permutations, which permute the particles inside their clusters. }
\end{proposition}

\noindent It means that the relations \ 
\begin{equation}
x_{i}=y_{i},  \label{x_i=y_i}
\end{equation}%
\begin{equation}
x_{i}-x_{\sigma _{i}}=i-\sigma _{i}.  \label{x_i-x_s_i=i-s_i}
\end{equation}%
hold for any $i=1,\ldots ,P$ .

We should remark that in the case of the continuous time ASEP, which
corresponds to the particular limit of our model with $\lambda \rightarrow 0$%
, \cite{schutz}, the contribution to the continuous time version of (\ref%
{<x|x'>}) comes from the only summand, which corresponds to the identical
permutation $\sigma _{i}=i$. In the Appendix A, we prove the above
proposition for the general case under consideration, while the continuous
time case follows from the proof.

Given the configuration $X=Y$ with any pair of coordinates $x_{i}$, $%
x_{\sigma _{i}}$ satisfying (\ref{x_i=y_i},\ref{x_i-x_s_i=i-s_i}), we can
calculate the integral (\ref{integral1}), which is equal to $(-\lambda
)^{i-\sigma _{i}}$ for $i>\sigma _{i}$ and $1$ for $i\leq \sigma _{i}$. The
summation over all permutations of particles within the clusters leads to
the determinant of $P\times P$ matrix

\begin{equation}
\int dZ\left\langle X|B_{Z}\right\rangle \left\langle \overline{B}%
_{Z}|X\right\rangle =W(X)\det \mathbf{F}_{0},  \label{<x|x>}
\end{equation}%
which has a block diagonal form%
\begin{equation}
\mathbf{F}_{0}\mathbf{=}\left\vert 
\begin{array}{cccc}
\mathbf{S}(n_{1}) &  &  & \text{{\large 0}} \\ 
& \mathbf{S}(n_{2}) &  &  \\ 
&  & \mathbf{S}(n_{3}) &  \\ 
\text{{\large 0}} &  &  & \ddots%
\end{array}%
\right\vert .  \label{F_0}
\end{equation}%
Here, $\mathbf{S}(n)$ is the $n\times n$ matrix%
\begin{equation}
\mathbf{S}(n)=\left\vert 
\begin{array}{ccccc}
1 & -\lambda & \left( -\lambda \right) ^{2} & \left( -\lambda \right) ^{3} & 
\ldots \\ 
1 & 1 & -\lambda & \left( -\lambda \right) ^{2} & \ldots \\ 
1 & 1 & 1 & -\lambda & \ldots \\ 
1 & 1 & 1 & 1 & \ldots \\ 
\ldots & \ldots & \ldots & \ldots & \ldots%
\end{array}%
\right\vert ,  \label{S(n)=det}
\end{equation}%
which corresponds to a cluster consisting of $n$ particles. It can be
transformed to the triangular form by subtracting neighboring columns, which
allows the calculation of its determinant: 
\begin{equation}
\det \mathbf{S}(n)=\left( 1+\lambda \right) ^{n-1}=\left( 1-v\right) ^{-n+1}
\end{equation}%
The product of such terms over all clusters yields just the value inverse to
the weight $W(X)$ defined above: 
\begin{equation}
\det \mathbf{F}_{0}=\left( 1-v\right) ^{\mathcal{N}_{c}\left( X\right)
-P}=1/W(X)  \label{det F_0}
\end{equation}%
The substitution to the formula (\ref{<x|x>}) completes the proof of the
relation (\ref{<x|x'>}). We should note that the matrix (\ref{S(n)=det}) has
a lower triangular form in the limit $\lambda \rightarrow 0,$ with all the
units on the main diagonal, as should be in the continuous time limit.

The integral (\ref{f(a,b,t=INT)}) can be evaluated in terms of the
hypergeometric functions. The sum in the r.h.s. of the formula (\ref%
{<x|T^t|x_0>-explicit form}) can be rewiten in the determinant form 
\begin{equation}
\left\langle X\right\vert \mathbf{T}^{t}\left\vert X^{0}\right\rangle
=W(X)\det \left[ f(x_{i}-x_{j}^{0},i-j,t)\right] _{i,j=1,\ldots ,M},
\end{equation}%
while as follows from (\ref{det F_0}) the factor $W(X)$ is the inverse of
the determinant 
\begin{equation}
W(X)=1/\det \left[ f(x_{i}-x_{j},i-j,0)\right] _{i,j=1,\ldots ,M}
\label{1/W(X)}
\end{equation}%
The resulting expression is given in the Section 2, (\ref{P(X,t|X_0,0)}-\ref%
{f(a,b,t)}).

Once the answer has been obtained, one can directly reexamine it. First, we
note that the correct initial conditions 
\begin{equation}
\frac{\det \mathbf{F}(X,X^{0},0)}{\det \mathbf{F}(X,X,0)}=\delta (X,X^{0})
\end{equation}%
follow from the Proposition 1. We should note that unlike the continuous\
time case, the contribution to the determinants comes not only from the
elements of the main diagonal of the matrix $\mathbf{F}(X,X,0)$, but from
the blocks connecting the particles within the same clusters. Then the
equality 
\begin{equation}
\frac{\det \mathbf{F}(X,X^{0},t+1)}{\det \mathbf{F}(X,X,0)}=\sum_{\left\{
X^{\prime }\right\} }T(X,X^{\prime })\frac{\det \mathbf{F}(X^{\prime
},X^{0},t)}{\det \mathbf{F}(X^{\prime },X^{\prime },0)}  \label{master1}
\end{equation}%
can be proved in spirit of the proof given by Sch\"{u}tz \cite{schutz}. We
first use just established fact (\ref{1/W(X)}) that the determinants in the
denominators in both sides are the inverse stationary weights $1/W(X)$ and $%
1/W(X^{\prime })$. Due to (\ref{T=T W(x')/W(x)}), brought to r.h.s., they
transform the transition probability $T(X,X^{\prime })$ to $%
T_{0}(X,X^{\prime })$%
\begin{equation}
\det \mathbf{F}(X,X^{0},t+1)=\sum_{\left\{ X^{\prime }\right\}
}T_{0}(X,X^{\prime })\det \mathbf{F}(X^{\prime },X^{0},t).  \label{master2}
\end{equation}%
Now one should use the properties of the matrix elements $%
f(x_{i}-x_{j}^{0},i-j,t)$ to show that the equality (\ref{master2}) holds.
The solution being the linear combination of the eigenvectors, all the
properties of its constituents follow from the corresponding properties of
the Bethe ansatz. Particularly the constraint (\ref{constraint}) imply that
the function $f(a,b,t)$ obeys the relation%
\begin{eqnarray}
0 &=&\left[ \lambda f\left( q_{1},n_{1}-1,t\right) +f\left(
q_{1},n_{1},t\right) \right] \left[ f\left( q_{2},n_{2},t\right) -f\left(
q_{2},n_{2}+1,t\right) \right]   \notag \\
&&-\left[ \lambda f\left( q_{2}-1,n_{2}-1,t\right) +f\left(
q_{2}-1,n_{2},t\right) \right] \left[ f\left( q_{1}+1,n_{1},t\right)
-f\left( q_{1}+1,n_{1}+1,t\right) \right] ,  \label{f-constraint}
\end{eqnarray}%
where $n_{1},n_{2}$ are arbitrary integers, $q_{1}=1-P,\ldots ,P-2$ and $%
q_{2}=2-P,\ldots ,P-1$. \ As above this relation ensures that in the
physical domain the master equation can be written in form of the equation
for noniteracting particles. The latter is satisfied if every individual
function $f(x_{i}-x_{j}^{0},i-j,t)$ for any pair $i,j$ satisfies the
equation for a free single particle%
\begin{equation}
f(q,n,t+1)=vpf(q,n-1,t)+\left( 1-v\right) f(q,n,t).  \label{f-free}
\end{equation}%
Both relations can be checked by substitution of the explicit form of the
function $f(a,b,t)$ and follow from the contiguous relations for
hypergeometric functions. Thus, the expression obtained is indeed the
solution of the master equation.

We should note that if one would use the resulting determinant form as an
ansatz with an unknown function $f(a,b,t)$, the relations (\ref{f-constraint}%
,\ref{f-free}) can be considered as difference equations, which are to be
solved to find this function without the above Bethe ansatz analysis. Such a
representation can be considered as an alternative treatment of the Bethe
ansatz. Of course it leads to the same final result. A problem, however,
occurs on the way. To solve the difference equations one should impose the
initial conditions, e.g. $f(1-P,n,0)$, which reproduce the correct value of
the resulting determinant. Up to our knowledge there is no a regular
procedure of doing this. In fact we encountered the equivalent problem in
the above Bethe ansatz analysis, which is how to choose the measure of
integration over $Z$. Luckily, the simplest choice made yields the correct
result. We, however, can avoid considering this problem at all by using the
constructive approach to the probability derivation based of the direct
enumeration of particle trajectories. This approach is presented in the next
section.

\section{Geometric approach.}

In this section, we give an alternative solution of the ASEP with the
parallel update based on a combinatorial analysis of particle trajectories.
We introduce a geometrical treatment considering sequentially three models
of interacting particles with increasing complexity of interaction. The
simplest one is the free-fermion model of non-intersecting trajectories,
known in physical literature as the problem of "vicious walkers" \cite%
{Fisher} and in mathematical literature as the Gessel-Viennot theorem \cite%
{Gessel}. Then, we consider the ASEP with the backward sequential update to
demonstrate how the Bethe ansatz appears with a complication of interaction.
Finally, we apply the developed approach to our main problem of the ASEP
with the parallel update.

\subsection{Combinatorial ansatz for vicious walkers}

Consider $P$ particles labelled $\{1,2,\ldots,P\}$ hopping in one direction
on an infinite one-dimensional lattice. The discrete space-time dynamics of
their motion can be described as a set of continuous broken trajectories on
a triangle lattice $\Lambda$, which is obtained from the square lattice by
adding a diagonal bond between the upper left corner and the lower right
corner of each elementary square. Let $(x,t)$ be integer space-time
coordinates of a particle on $\Lambda$, where the vertical time axis is
directed down and the horizontal space axis is directed right. A trajectory
of each particle is a sequence of connected vertical and diagonal bonds of $%
\Lambda$. The diagonal bonds correspond to jumps of particles to their right
for a unit time-step with probability $v$. The vertical bonds correspond to
stays of particles at fixed sites during the unit time-step with probability 
$1-v$. The initial positions of particles are $x^0_1 < x^0_2 < \ldots <
x^0_P $. Let our particles be found at sites $x_1 < x_2 < \ldots < x_P$ by
the moment of time $t$. The problem of vicious walkers is to find the
conditional probability for the particles to reach the positions $%
X=(x_1,x_2,\ldots,x_P); x_1 < x_2 < \ldots < x_P$ from the initial positions 
$X^0=(x^0_1,x^0_2,\ldots,x^0_P)$ for time $t$ so that no pairs of
trajectories are intersected during time $t$. In other words, one assumes
that every site can be occupied by at most one particle at every moment of
discrete time and if this rule is violated at a moment $t^{\prime}<t$, the
process stops.

We start with consideration of the one-particle motion on the infinite
lattice. Let $\mathcal{T}_{t}(x|x^{0})$ be a set of one-particle
trajectories, which are starting at $(x^{0},0)$ and finishing at $(x,t)$.
Each trajectory $q\in \mathcal{T}_{t}(x|x^{0})$ is realized with probability 
$v^{x-x^{0}}(1-v)^{t-x+x^{0}}$. Therefore, the total probability for the
particle to reach $x$ from $x^{0}$ for time $t$ is 
\begin{equation}
P(x,t|x^{0},0)=v^{x-x^{0}}(1-v)^{t-x+x^{0}}\Vert \mathcal{T}%
_{t}(x|x^{0})\Vert =F_{0}(x-x^{0}|t).  \label{OneParticleProb}
\end{equation}%
where 
\begin{equation}
F_{0}(x|t)={\binom{t}{x}}v^{x}(1-v)^{t-x}  \label{F}
\end{equation}%
is the weight of one-particle trajectories in the set $\mathcal{T}%
_{t}(x|x^{0})$. For the case of $P$ particles, the set ${}\mathbb{S}_{P}$ of
all possible free trajectories is a direct product of one-particle sets of
trajectories reaching $(X,t)$ from $(X^{0},0)$: 
\begin{equation}
{}\mathbb{S}_{P}=\mathcal{T}_{t}(x_{1}|x_{1}^{0})\otimes \mathcal{T}%
_{t}(x_{2}|x_{2}^{0})\otimes \ldots \otimes \mathcal{T}_{t}(x_{P}|x_{P}^{0})
\end{equation}%
The set $\mathbb{S}_{P}$ contains non-intersecting and intersecting
trajectories. The latter cases are forbidden by the single-occupation rule
and should be subtracted in evaluations of probability $P(X,t|X^{0},0)$.

Consider first the case $P=2$. We denote the set of forbidden elements by $%
\mathbb{U}_{12}$, $\mathbb{U}_{12}\subset \mathbb{S}_{2}$ where indices $1,2$
emphasize that the order of final points of intersecting trajectories
coincides with the initial one. To cancel the contribution of forbidden
elements, we introduce an auxiliary set of pairs of trajectories 
\begin{equation}
\mathbb{A}_{21}=\mathcal{T}_{t}(x_{2}|x_{1}^{0})\otimes \mathcal{T}%
_{t}(x_{1}|x_{2}^{0}),
\end{equation}%
where the final coordinates are permuted with respect to those in ${}\mathbb{%
S}_{2}$. It is easy to see that all elements of the set $\mathbb{A}_{21}$
are pairs of crossing trajectories. Each intersecting pair $q\in \mathbb{S}%
_{2}$ has a first collision point $(x_{c},t_{c})$, i.e. the space-time
point, where the trajectories meet for a first time. Consider the set ${}%
\mathbb{S}_{2}(x_{c},t_{c})$ of intersecting trajectories with the fixed
first collision point $(x_{c},t_{c})$. The sets ${}\mathbb{S}%
_{2}(x_{c},t_{c}),\;(x_{c},t_{c})\in \Lambda $ break the set $\mathbb{U}%
_{12} $ into subsets parameterized by coordinates $(x_{c},t_{c})$. For all $%
(x_{c}^{\prime },t_{c}^{\prime })\neq (x_{c}^{\prime \prime },t_{c}^{\prime
\prime })$ we have 
\begin{equation}
{}\mathbb{S}_{2}(x_{c}^{\prime },t_{c}^{\prime })\cap {}\mathbb{S}%
_{2}(x_{c}^{\prime \prime },t_{c}^{\prime \prime })=\emptyset
\end{equation}%
and 
\begin{equation}
\bigcup_{(x_{c},t_{c})\in \Lambda }\mathbb{S}_{2}(x_{c},t_{c})=\mathbb{U}%
_{12}  \label{U12}
\end{equation}

For every set ${}\mathbb{S}_{2}(x_{c},t_{c})$, there exists a uniquely
defined set $\mathbb{A}_{21}(x_{c},t_{c})\subset \mathbb{A}_{21}$ obtained
from ${}\mathbb{S}_{2}(x_{c},t_{c})$ by permutation of tails of all
trajectories beginning at the first collision point. The sets ${}\mathbb{S}%
_{2}(x_{c},t_{c})$ and ${}\mathbb{A}_{21}(x_{c},t_{c})$ are geometrically
equivalent,as there is one-to-one correspondence between their elements.
Therefore $\Vert {}\mathbb{S}_{2}(x_{c},t_{c})\Vert =\Vert {}\mathbb{A}%
_{2}(x_{c},t_{c})\Vert $. Like the sets ${}\mathbb{S}_{2}(x_{c},t_{c})$, the
sets ${}\mathbb{A}_{21}(x_{c},t_{c}),\;(x_{c},t_{c})\in \Lambda $ break the
set ${}\mathbb{A}_{21}$ into subsets: for all $(x_{c}^{\prime
},t_{c}^{\prime })\neq (x_{c}^{\prime \prime },t_{c}^{\prime \prime })$ we
have 
\begin{equation}
\mathbb{A}_{21}(x_{c}^{\prime },t_{c}^{\prime })\cap \mathbb{A}%
_{21}(x_{c}^{\prime \prime },t_{c}^{\prime \prime })=\emptyset  \label{empty}
\end{equation}%
and 
\begin{equation}
\bigcup_{(x_{c},t_{c})\in \Lambda }\mathbb{A}_{21}(x_{c},t_{c})=\mathbb{A}%
_{21}.  \label{A21}
\end{equation}

It follows from (\ref{U12},\ref{empty},\ref{A21}) that the whole sets $%
\mathbb{U}_{12}$ and ${}\mathbb{A}_{21}$ are geometrically equivalent and $%
\Vert \mathbb{U}_{12}\Vert =\Vert \mathbb{A}_{21}\Vert $. Given $(X,t)$ and $%
(X^{0},0)$, all elements of ${}\mathbb{S}_{2}$ have the same weight 
\begin{equation*}
Q=\prod_{i=1}^{2}v^{x_{i}-x_{i}^{0}}(1-v)^{t-x_{i}+x_{i}^{0}}.
\end{equation*}%
Then, we have for the probability 
\begin{eqnarray}
P(X,t|X^{0},0) &=&Q\times \left( \Vert \mathbb{S}_{2}\Vert -\Vert \mathbb{A}%
_{21}\Vert \right) = \\
&=&\left(
F_{0}(x_{1}-x_{1}^{0}|t)F_{0}(x_{2}-x_{2}^{0}|t)-F_{0}(x_{2}-x_{1}^{0}|t)F_{0} (x_{1}-x_{2}^{0}|t)\right) =
\notag \\
&=&\det\mathbf{M},  \notag
\end{eqnarray}%
where the elements of the $2\times 2$ matrix $\mathbf{M}$ are 
\begin{equation}
M_{i,j}=F_{0}(x_{j}-x_{i}^{0}|t),\hspace{0.5cm}i,j=1,2.
\end{equation}

This result may be easily generalized for the $P$-particle system of vicious
walkers \cite{Fisher}. The set $\mathbb{S}_{P}$ contains the subset of
intersecting trajectories ${}\mathbb{U}_{12...P}$. We introduce $P!-1$
auxiliary sets 
\begin{equation}
{}\mathbb{A}_{\sigma }=\mathcal{T}_{t}(x_{\sigma _{1}}|x_{1}^{0})\otimes 
\mathcal{T}_{t}(x_{\sigma _{2}}|x_{2}^{0})\otimes \ldots \otimes \mathcal{T}%
_{t}(x_{\sigma _{P}}|x_{P}^{0})
\end{equation}%
where $\sigma =\{\sigma _{1},\ldots ,\sigma _{P}\}$ is any permutation of
numbers $1,2,...,P$ beside the identical one. Each element of the set 
\begin{equation}
{}\mathbb{U}_{12...P}\bigcup \{\bigcup_{\sigma \neq 1}\mathbb{A}_{\sigma }\}
\label{UA}
\end{equation}%
containing trajectories intersecting at given point $(x_{c},t_{c})$, has a
unique geometrically identical counterpart with a pair of permuted indices,
e.g. $12...i...j...P$ and $12...j...i...P$. Introducing the sign of
permutation $(-1)^{\sigma }$, we can ascribe opposite signs to two
counterparts. Then, we obtain the known result \cite{Fisher}: 
\begin{eqnarray}
P(X,t|X^{0},0) &=&Q\times \left( \Vert \mathbb{S}_{P}\Vert +\sum_{\sigma
\neq 1}(-1)^{\sigma }\Vert \mathbb{A}_{\sigma }\Vert \right) = \\
&=&\det \mathbf{M},  \notag  \label{det1}
\end{eqnarray}%
where 
\begin{equation}
M_{i,j}=F_{0}(x_{j}-x_{i}^{0}|t),\qquad i,j=1,2,...P.  \label{Mij}
\end{equation}

\subsection{Combinatorial treatment of Bethe ansatz}

The vicious walkers are locally interacting particles. In this section, we
consider the system of particles with the interaction of a non-zero range,
which takes place in the case of the totally asymmetric exclusion process
with the backward sequential update \cite{open case 2}. Consider again $P$
particles hopping in one direction on an infinite one-dimensional lattice.
The interaction between particles can be defined by the following rules:

1. Trajectories of particles do not intersect (every site can be occupied by
at most one particle).

2. A particle stays at its own site with probability $1$ if the target site
is occupied by another particle during the step of discrete time.

Like the previous case, the trajectory of each particle is a sequence of
vertical and diagonal bonds on the lattice $\Lambda $. Each diagonal bond
has weight $v$ and the vertical bond weight $1-v$. In view of more
complicated interaction, it is convenient to decompose the set of all free
trajectories into more detailed subsets. For each vertical bond $%
[(x,t^{\prime }),(x,t^{\prime }+1)]$, the trajectory of $i$-th particle
passing this bond can be decomposed into two parts $[(x_{i}^{0},0)%
\rightarrow (x,t^{\prime })|1|(x,t^{\prime }+1)\rightarrow (x_{i},t)]$ and $%
[(x_{i}^{0},0)\rightarrow (x,t^{\prime })|-v|(x,t^{\prime }+1)\rightarrow
(x_{i},t)]$ where the value between vertical bars means the weight of the
bond $[(x,t^{\prime }),(x,t^{\prime }+1)]$. These new trajectories are
geometrically equivalent, but the first trajectory passes the selected
vertical bond with weight $1$ and second one with weight $-v$. We make this
decomposition for each vertical bond of each trajectory and denote the whole
set of decomposed trajectories of $i$-th particle by $\mathcal{T}%
_{t}(x_{i}|x_{i}^{0})$ using the same notation as for the set of
one-particle trajectories in the previous section.

The weight of the trajectory $q\in \mathcal{T}_{t}(x_{i}|x_{i}^{0})$ is a
product of the weights of its bonds. 
\begin{equation}
\mu (q)=\prod_{k=1}^{t}\mu (k\mathtt{\small -}\mathrm{th\quad bond\quad
of\quad }q)
\end{equation}%
The weight of the set of trajectories is a sum of weights of its elements.
In accordance with the above definition (\ref{OneParticleProb}), the weight
of the whole set $\mathcal{T}_{t}(x_{i}|x_{i}^{0})$ is 
\begin{equation}
\mu (\mathcal{T}_{t}(x_{i}|x_{i}^{0}))=F_{0}(x_{i}-x_{i}^{0}|t)
\end{equation}

The set of free $P$-particle trajectories $\mathbb{S}_{P}$,

\begin{equation}
\mathbb{S}_{P}=\mathcal{T}_{t}(x_{1}|x_{1}^{0})\otimes \mathcal{T}%
_{t}(x_{2}|x_{2}^{0})\otimes \ldots \otimes \mathcal{T}_{t}(x_{P}|x_{P}^{0})
\label{SP}
\end{equation}%
has the weight

\begin{equation}
\mu (\mathbb{S}_{P})=\sum_{q\in \mathbb{S}_{P}}\mu
(q)=\prod_{i=1}^{P}v^{x_{i}-x_{i}^{0}}(1-v)^{t-x_{i}+x_{i}^{0}}{\binom{t}{%
x_{i}-x_{i}^{0}}}.
\end{equation}

As in the case of vicious walkers, the set $\mathbb{S}_{P}$ contains a
subset of unallowed elements,${}\mathbb{U}_{12...P}$, which should be
excluded. By the rules 1 and 2, an element of $\mathbb{S}_{P}$ is unallowed,
if there is at least one pair of intersecting trajectories, or if there are
two neighboring vertical bonds from which the left one has weight $-v$. To
cancel unallowed elements we start as above with the case $P=2$ and
construct an auxiliary set of trajectories $\mathbb{\ A}_{21}$. We will see
that, due to nonlocality, the auxiliary set has more complicated structure

\begin{equation}
\mathbb{\ A}_{21}=\bigcup_{k_1=0}^{\infty}\bigcup_{k_2=0}^{1} \mathcal{T}%
_t(x_2|x_1^0-k_1) \otimes \mathcal{T}_t(x_1|x_2^0-k_2).  \label{defA21}
\end{equation}

Beginning the construction, we notice that every unallowed pair has a first
collision point $(x_c,t_c)\in \Lambda$, where the particles come for the
first time to neighboring sites at a moment $t_c<t$. The first trajectory
reaches the site $(x_c,t_c)$ from $(x_1^0,0)$ and second one reaches the
site $(x_c+1,t_c)$ from $(x_2^0,0)$ and then it makes the vertical step to
the site $(x_c+1,t_c+1)$.

The first trajectory has, just after the collision, a diagonal bond $%
[(x_c,t_c),(x_c+1,t_c+1)]$ with weight $v$ (refered to as a collision of the
first type) or vertical bond $[(x_c,t_c),(x_c,t_c+1)]$ with weight $-v$
(refered to as the collision of the second type). Notice, that trajectories,
which have the vertical bond $[(x_c,t_c),(x_c,t_c+1)]$ with weight $1$ are
allowed. Thus, we have two types of unallowed elements of $\mathbb{\ U}_{12}$
for fixed $(x_c,t_c)$. Denote these subsets by $\mathbb{\ V}(x_c,t_c)$ for
the first type and $\mathbb{\ W}(x_c,t_c)$ for the second one. For all $%
(x_c^{\prime},t_c^{\prime})\neq (x_c^{\prime\prime},t_c^{\prime\prime})$ we
have 
\begin{eqnarray}
\mathbb{\ V}(x_c^{\prime},t_c^{\prime})\cap \mathbb{\ V}(x_c^{\prime%
\prime},t_c^{\prime\prime})=\emptyset,  \notag \\
\mathbb{\ W}(x_c^{\prime},t_c^{\prime}) \cap \mathbb{\ W}(x_c^{\prime%
\prime},t_c^{\prime\prime})=\emptyset,  \notag \\
\mathbb{\ V}(x_c^{\prime},t_c^{\prime}) \cap \mathbb{\ W}(x_c^{\prime%
\prime},t_c^{\prime\prime})=\emptyset,  \notag
\end{eqnarray}
and 
\begin{equation}
\bigcup_{(x_c,t_c)\in \Lambda} \left( \mathbb{\ V}(x_c,t_c) \cup \mathbb{\ W}%
(x_c,t_c) \right)=\mathbb{U}_{12}  \label{VandW}
\end{equation}

For each space-time point $(x_c,t_c)\in \Lambda$, we construct the sequence
of pairs of trajectories $\mathbb{\ A}_{v}(k_1,k_2)$ and $\mathbb{\ A}%
_{w}(k_1,k_2)$, ($k_1=0,1,2,\ldots$ and $k_2=0,1$,arguments $x_c,t_c$ are
omitted), obtained from $\mathbb{\ V}$ and $\mathbb{\ W}$ by permutation of
tails in each pair of trajectories and shifting the initial parts in
negative direction.

Explicit form of transformations $\mathbb{\ V} \Rightarrow \mathbb{\ A}%
_{v}(k_1,k_2)$ and $\mathbb{\ W}\Rightarrow \mathbb{\ A}_{w}(k_{1},k_{2})$
are given in Appendix B. Using these relations, we can express the
probability 
\begin{equation}
P(X,t|X^{0},0)=\mu (\mathbb{\ S}_{2}\setminus \mathbb{\ U}_{12})  \label{PT2}
\end{equation}%
via the weight of all unallowed configurations 
\begin{equation}
\mu (\mathbb{\ U}_{12})=\sum_{(x_{c},t_{c})}\sum_{k=0}^{\infty }\left( \mu (%
\mathbb{\ A}_{v}(k,0))+\mu (\mathbb{\ A}_{w}(k,0))-\mu (\mathbb{\ A}%
_{v}(k,1))-\mu (\mathbb{\ A}_{w}(k,1))\right) .  \label{muU12}
\end{equation}

Our next aim is to bring this expression to a determinant form similar to (%
\ref{det1}). Consider operator $\hat{a}_i$, which acts on the set of the
free trajectories of $i$-th particle and gives the set of free trajectories
with the shifted origin by one lattice space to the left: 
\begin{equation}
\hat{a}_i \mathcal{T}_t(x_j|x_i^0)=\mathcal{T}_t(x_j|x_i^0-1)
\end{equation}

The operator representation allows us to write (\ref{muU12}) in compact
form: 
\begin{equation}
\mu(\mathbb{\ U}_{12})=\mu\left(\frac{1-\hat{a}_2}{1-\hat{a}_1}\mathcal{T}%
_t(x_2|x_1^0) \otimes\mathcal{T}_t(x_1|x_2^0)\right),  \label{oper}
\end{equation}
where the denominator is defined by its expansion 
\begin{equation}
\frac{1}{1-\hat{a}_1}=\sum_{k=0}^{\infty}\hat{a}_1^k.
\end{equation}
and the summation of operators means the joining of the sets.

The crucial property of the operator expression in (\ref{oper}) is its
factorization with respect to indices 1 and 2. Introducing the functions 
\begin{equation}
F_{1}(x_{2}-x_{1}^{0}|t)=\mu \left( (1-\hat{a}_{1})^{-1}\mathcal{T}%
_{t}(x_{2}|x_{1}^{0})\right) =\sum_{k=0}^{\infty }F_{0}(x_{2}-x_{1}^{0}+k|t),
\label{F1}
\end{equation}%
and 
\begin{equation}
F_{-1}(x_{1}-x_{2}^{0}|t)=\mu \left( (1-\hat{a}_{2})\mathcal{T}%
_{t}(x_{1}|x_{2}^{0})\right) =\sum_{k=0}^{1}F_{0}(x_{1}-x_{2}^{0}+k|t)
\label{F-1}
\end{equation}%
we obtain the two-particle probability $P(X,t|X^{0},0)=\mu (\mathbb{\ S}%
_{2})-\mu (\mathbb{\ U}_{12})$ in the form 
\begin{equation}
P(X,t|X^{0},0)=F_{0}(x_{1}-x_{1}^{0}|t)F_{0}(x_{2}-x_{2}^{0}|t)-F_{1}(x_{2}-x_{1}^{0}|t) F_{-1}(x_{1}-x_{2}^{0}|t)
\label{prob2}
\end{equation}%
or 
\begin{equation}
P(X,t|X^{0},0)=\det\;\mathbf{M},  \label{det2}
\end{equation}%
where 
\begin{equation}
M_{i,j}=F_{i-j}(x_{i}-x_{j}^{0}|t) \qquad i,j=1,2
\end{equation}

To generalize the determinant formula (\ref{det2}) to the general P-particle
case, we should take into consideration several additional properties of
intersecting trajectories. First,we may organize the procedure of exclusion
of unallowed elements of $\mathbb{\ U}_{12...P}$ in an ordered way. Starting
with the top line of the lattice $\Lambda $, we examine all free
trajectories of the set $\mathbb{\ S}_{P}$, row by row, until we meet the
first collision point where unallowed trajectories are cancelled with
elements of the auxiliary set $\mathbb{\ A}_{\sigma }$.

If the number of particles $P>2$, the elementary squares associated with the
collision point $%
(x_{c},t_{c}),(x_{c},t_{c})(x_{c}+1,t_{c})(x_{c},t_{c}+1)(x_{c}+1,t_{c}+1),$
may occur several times in one horizontal strip of $\Lambda $. If squares
filled by interacting trajectories are separated one from another by a gap
of empty sites, the above arguments can be applied to each pair of
interacting trajectories independently. The crucial case for the Bethe
ansatz is a situation when the elementary squares are nearest neighbors. The
specific property of the totally ASEP is that, in each pair of interacting
trajectories, the right trajectory remains free and interacts with the next
trajectory independently on its left neighbors. Therefore, we can analyse
the interaction between particles considering successively elementary
squares in each row from left to right starting from an arbitrary empty
square until all unwanted trajectories will be removed.

After $t^{\prime}$ steps from the top to bottom, one obtains the set of path
configurations $q \in \mathbb{\ S}_{P}$ which are allowed in the first $%
t^{\prime}$ rows, and the set of auxiliary configurations yet not involved
into cancellation procedures. Remembering that all elements of $\mathbb{\ A}%
_{\sigma}$ for all $\sigma\neq 1$ have end points permuted with respect to
the original order $12...P$, we conclude that each element of $\mathbb{\ A}%
_{\sigma}$ contains at least one collision point. Therefore, all elements of
the auxiliary set will be cancelled after $t^{\prime}=t$ steps with
unallowed elements of $\mathbb{\ U}_{12...P}$.

The described way of exclusion of unallowed configurations implies a
successive construction of the auxiliary set $\mathbb{\ A}_{\sigma}$. We
notice that each two intersecting trajectories in $\mathbb{\ A}_{21}$ are
non-equivalent: one of them belongs to the particle which overtakes another
and can be called "active". On the contrary, the second particle is
"passive". In the case $P>2$, one trajectory can overtake $m$ others, and we
call it "m-active". Similarly, the "m-passive" trajectories appear.

Assume, that the trajectory of a given particle has $m$ active
intersections. It means that it participates $m$ times in the cancellation
procedure and its starting point is shifted $m$ times to arbitrary distances
in the negative direction. As a result, the auxiliary set associated with
the free trajectory between points $x_i^0$ and $x_j$ becomes 
\begin{equation}
\frac{1}{(1-\hat{a}_i)^m}\mathcal{T}_t(x_j|x_i^0)
\end{equation}
Similarly, for trajectories having $m$ passive intersections we get 
\begin{equation}
(1-\hat{a}_i)^m\mathcal{T}_t(x_j|x_i^0)
\end{equation}

The weights of sets of m-active and m-passive trajectories are given by
functions introduced in \cite{schutz}: 
\begin{equation}
F_m(x_j-x_i^0|t)= \mu \left(\frac{1}{(1-\hat{a}_i)^m}\mathcal{T}%
_t(x_j|x_i^0)\right)= \sum_{k=0}^{\infty}\left( 
\begin{array}{c}
k+m-1 \\ 
m-1%
\end{array}
\right) F_0(x_j-x_i^0+k|t) ,  \label{Fplus}
\end{equation}
and 
\begin{equation}
F_{-m}(x_j-x_i^0|t)= \mu \left((1-\hat{a}_i)^m \mathcal{T}%
_t(x_j|x_i^0)\right)= \sum_{k=0}^{m}\left( 
\begin{array}{c}
m \\ 
k%
\end{array}
\right) (-1)^k F_0(x_j-x_i^0+k|t) ,  \label{Fminus}
\end{equation}

Activity $m$ of each trajectory is defined uniquely by the permutation of
numbers $\sigma $, so we have for the weight of auxiliary set $\mathbb{\ A}%
_{\sigma }$ 
\begin{equation}
\mu (\mathbb{\ A}_{\sigma })=\prod_{i=1}^{P}F_{\sigma _{i}-i}(x_{\sigma
_{i}}-x_{i}^{0}|t)
\end{equation}%
This product is a term of expansion of the determinant $\det\mathbf{M}$ with
matrix elements for $\sigma \neq 1$ 
\begin{equation}
M_{i,j}=F_{i-j}(x_{i}-x_{j}^{0}|t) \qquad i,j=1,2,...,P  \label{M1P}
\end{equation}%
It follows from (\ref{SP}) that the term corresponding to the identical
permutation $\sigma =1$ is 
\begin{equation}
\mu (\mathbb{\ S}_{P})=\prod_{i=1}^{P}F_{0}(x_{i}-x_{i}^{0}|t)
\end{equation}%
Collecting the contributions from the set $\mathbb{\ S}_{P}$ and all
auxiliary sets $\mathbb{\ A}_{\sigma }$, we obtain the determinant formula 
\cite{schutz}%
\begin{equation}
P(X,t|X^{0},0)=\det\mathbf{M},  \label{detP}
\end{equation}%
which is valid for all $P\geq 1$.

\subsection{Combinatorial solution for parallel update}

The TASEP with parallel update is characterized by the rules

1. Trajectories of particles do not intersect.

2. A particle stays at its own site with probability $1$ if the target site
is occupied by another particle just before the step of discrete time.

The first rule coincides with that for the the backward sequential update,
however the second rules are different. The combinatorial solution for
parallel update accumulates all main steps of two preceding sections. We
start with the construction of different types of unallowed elements of $%
\mathbb{\ U}_{12}$ at the fixed collision point $(x_c,t_c)$. Two of them, $%
\mathbb{\ V}(x_c,t_c)$ and $\mathbb{\ W}(x_c,t_c)$ are the same as for the
sequential update. Two new sets $\mathbb{\ X}(x_c,t_c)$ and $\mathbb{\ Y}%
(x_c,t_c)$ are

$\mathbb{\ X}(x_{c},t_{c})$: the first trajectory has, just after the
collision, the diagonal bond $[(x_{c},t_{c}),(x_{c}+1,t_{c}+1)]$ with weight 
$v$ and the second one has also the diagonal bond $%
[(x_{c}+1,t_{c}),(x_{c}+2,t_{c}+1)]$ with weight $v$;

$\mathbb{\ Y}(x_{c},t_{c})$: the first trajectory has, after the collision,
the vertical bond $[(x_{c},t_{c}),(x_{c},t_{c}+1)]$ with weight $-v$ and the
second one has the diagonal bond $[(x_{c}+1,t_{c}),(x_{c}+2,t_{c}+1)]$ with
weight $v$.

Like (\ref{VandW}) we have 
\begin{equation}
\bigcup_{(x_c,t_c)\in \Lambda} \left( \mathbb{\ V}(x_c,t_c) \cup \mathbb{\ W}%
(x_c,t_c) \cup \mathbb{\ X}(x_c,t_c)\cup \mathbb{\ Y}(x_c,t_c) \right)=%
\mathbb{U}_{12}  \label{VWXY}
\end{equation}
Similarly, we introduce along with the auxiliary sets $\mathbb{\ A}%
_{v}(k_1,k_2)$ and $\mathbb{\ A}_{w}(k_1,k_2)$ new sets $\mathbb{\ A}%
_{x}(k_1,k_2)$ and $\mathbb{\ A}_{y}(k_1,k_2)$ obtained from $\mathbb{\ X}$
and $\mathbb{\ Y}$ by permutation of tails in each pair of trajectories and
shifting the initial parts by integers $k_1$ and $k_2$. Explicitly, the
transformations $\mathbb{\ X} \Rightarrow \mathbb{\ A}_{x}(k_1,k_2)$ and $%
\mathbb{\ Y}\Rightarrow \mathbb{\ A}_{y}(k_{1},k_{2})$ are given in Appendix
B.

\begin{figure}[t]
\unitlength=1mm \makebox(110,45)[cc] {\psfig{file=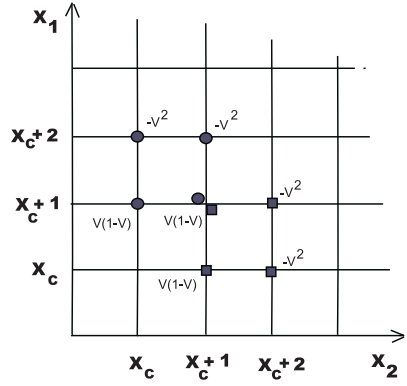,width=60mm}}
\caption{The reference points and the weights of the sets $\mathbb{V},%
\mathbb{{}W},\mathbb{{}X},\mathbb{{}Y}$ (full circles) and ${}\mathbb{A}%
_{v}(0,0),{}\mathbb{A}_{w}(0,0),{}\mathbb{A}_{x}(0,0),{}\mathbb{A}_{w}(0,0)$
(full squares)}
\label{diagram}
\end{figure}

The whole auxiliary set $\mathbb{\ A}_{21}$ has, instead of (\ref{normA21}),
a more complicated structure. To reveal it, we show schematically in Fig.(%
\ref{diagram}) the positions of sets $\mathbb{\ V},\mathbb{\ W},\mathbb{\ X},%
\mathbb{\ Y}$ and $\mathbb{\ A}_{v}(0,0),\mathbb{\ A}_{w}(0,0),\mathbb{\ A}%
_{x}(0,0),\mathbb{\ A}_{w}(0,0)$. It is easy to see that all these sets are
characterized by reference points from where trajectories reach the end
points $x_{1}$ and $x_{2}$ after the collision. For the set $\mathbb{\ V}$,
the point $x_{1}$ is reached from $x_{c}+1$, the point $x_{2}$ from $x_{c}+1$%
, for the set $\mathbb{\ W}$, $x_{1}$ is reached from $x_{c}$, $x_{2}$ from $%
x_{c}+1$; for the set $\mathbb{\ X}$, $x_{1}$ is reached from $x_{c}+1$, $%
x_{2}$ from $x_{c}+2$; and for the set $\mathbb{\ Y}$, $x_{1}$ is reached
from $x_{c}$, $x_{2}$ from $x_{c}+2$. The coordinates of the reference
points of $\mathbb{\ V},\mathbb{\ W},\mathbb{\ X},\mathbb{\ Y}$ are shown in
Fig.(\ref{diagram}) as full circles.

For the auxiliary sets $\mathbb{\ A}_{v}(0,0),\mathbb{\ A}_{w}(0,0),\mathbb{%
\ A}_{x}(0,0),\mathbb{\ A}_{w}(0,0)$, the coordinates of reference points
can be obtained by permutation of $x_{1}$ and $x_{2}$. The positions of
auxiliary sets are shown in Fig.(\ref{diagram}) as full squares.

The weights of sets in the "scattering zone" between $t_c$ and $t_c+1$ are: $%
v(1-v)$ for $\mathbb{\ V}$,$-v(1-v)$ for $\mathbb{\ W}$, $v^2$ for $\mathbb{%
\ X}$, $-v^2$ for $\mathbb{\ Y}$,$v(1-v)$ for $\mathbb{\ A}_{v}(0,0)$, $%
-v(1-v)$ for $\mathbb{\ A}_{w}(0,0)$, $v^2)$ for $\mathbb{\ A}_{x}(0,0)$ and 
$-v^2)$ for $\mathbb{\ A}_{y}(0,0)$.

Our aim is to combine the sets $\mathbb{\ A}_{v}(k_1,k_2),\mathbb{\ A}%
_{w}(k_1,k_2), \mathbb{\ A}_{x}(k_1,k_2),\mathbb{\ A}_{y}(k_1,k_2)$ in such
a way that all sets of unallowed trajectories are cancelled with their
auxiliary counterparts. Using the shift operators $\hat{a}_i$, we consider
sequentially four operator forms:

\begin{equation}
\hat{A}_1=\frac{1}{1+\lambda \hat{a}_2^{-1}},
\end{equation}

\begin{equation}
\hat{A}_{2}=\frac{1}{1-\hat{a}_{1}},
\end{equation}

\begin{equation}
\hat{A}_{3}=1+\lambda \hat{a}_{1}^{-1}
\end{equation}

\begin{equation}
\hat{A}_4=1-\hat{a}_2
\end{equation}
where $\lambda=v/(1-v)$.

\begin{figure}[t]
\unitlength=1mm \makebox(110,45)[cc] {\psfig{file=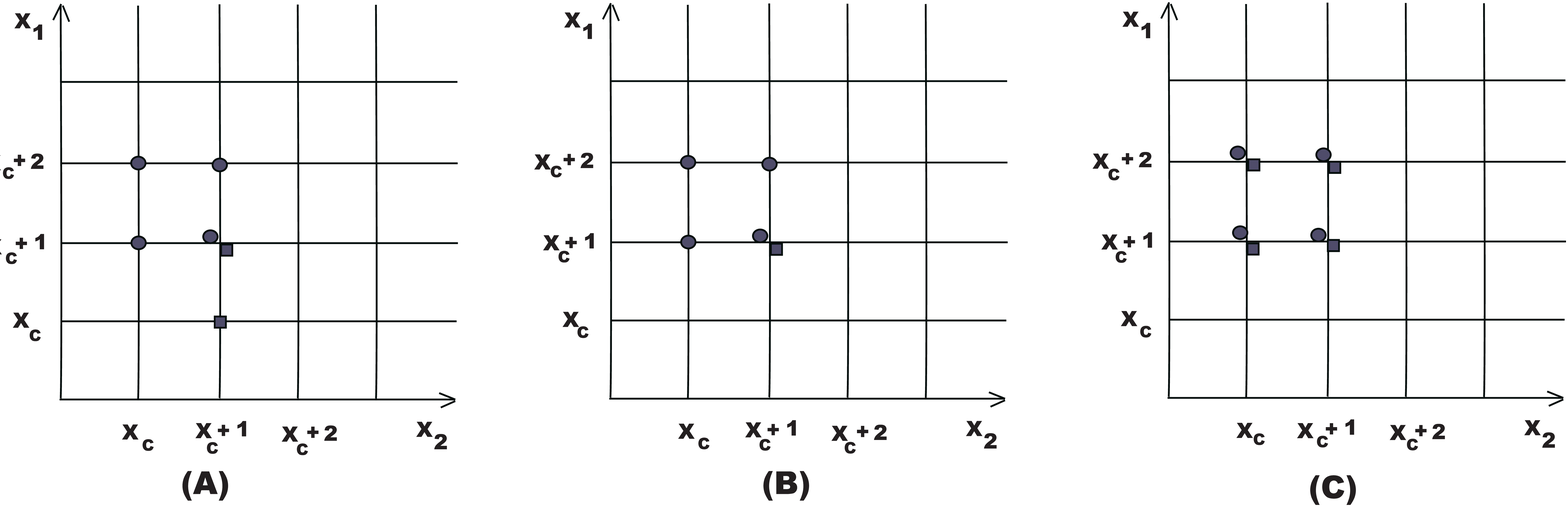,width=140mm}}
\caption{{The transformation of the reference points due to action of
operators $\hat{A}_1$ (a),$\hat{A}_2$ (b),and $\hat{A}_3 \hat{A}_4$ (c)}}
\label{diagram(abc)}
\end{figure}

Each shifted set contains trajectories where an initial part from $t=0$ to $%
t=t_{c}+1$ is geometrically equivalent to that of non-shifted trajectories.
It means that the reference points are also shifted under action of $\hat{a}%
_{i}$. As a result, the action of operator $\hat{A}_{1}$ leads to the
diagram Fig.(\ref{diagram(abc)}a), the action of operator $\hat{A}_{2}$
leads to the diagram Fig.(\ref{diagram(abc)}b),and the action of operator $%
\hat{A}_{3}\hat{A}_{4}$ gives the final diagram Fig.(\ref{diagram(abc)}c)
where positions of auxiliary sets coincide with those of unallowed sets $%
\mathbb{\ V},\mathbb{\ W},\mathbb{\ X},\mathbb{\ Y}$ together with their
weights. 

Thus, we may conclude that the whole set of auxiliary trajectories can be
obtained by action of operator $\hat{A}_{1}\hat{A}_{2}\hat{A}_{3}\hat{A}_{4}$%
, so we have 
\begin{equation}
\mu (\mathbb{U}_{12})=\mu \left( \frac{(1-\hat{a}_{2})}{(1-\hat{a}_{1})}%
\frac{(1+\lambda \hat{a}_{1}^{-1})}{(1+\lambda \hat{a}_{2}^{-1})}\mathcal{T}%
_{t}(x_{2}|x_{1}^{0})\otimes \mathcal{T}_{t}(x_{1}|x_{2}^{0})\right) ,
\label{operparal}
\end{equation}%
which should be compared with the analogous expression (\ref{oper}) for the
sequential update.

The operator expression (\ref{operparal}) is also factorizable with respect
to indices 1 and 2. It allows us to introduce functions which are
generalization of $F_{m}$ and $F_{-m}$ given by (\ref{Fplus},\ref{Fminus}): 
\begin{equation}
\tilde{F}_{m}(x_{j}-x_{i}^{0}|t)=\mu \left( \frac{(1+\lambda \hat{a}%
_{i}^{-1})^{m}}{(1-\hat{a}_{i})^{m}}\mathcal{T}_{t}(x_{j}|x_{i}^{0})\right)
\label{FF+oper}
\end{equation}%
and 
\begin{equation}
\tilde{F}_{-m}(x_{j}-x_{i}^{0}|t)=\mu \left( \frac{(1-\hat{a}_{i})^{m}}{%
(1+\lambda \hat{a}_{i}^{-1})^{m}}\mathcal{T}_{t}(x_{j}|x_{i}^{0})\right)
\label{FF-oper}
\end{equation}%
or, in the explicit form: 
\begin{equation}
\tilde{F}_{m}(N|t)=\sum_{n=0}^{m}\sum_{k=-n}^{\infty }\frac{m(m+k+n-1)!}{%
(k+n)!n!(m-n)!}\lambda ^{n}F_{0}(N+k|t),  \label{FFplus}
\end{equation}%
and 
\begin{equation}
\tilde{F}_{-m}(N|t)=\sum_{n=0}^{m}\sum_{k=-n}^{\infty }(-1)^{n}\frac{%
m(m+k+n-1)!}{(k+n)!n!(m-n)!}(-\lambda )^{k+n}F_{0}(N-k|t),  \label{FFminus}
\end{equation}%
The remainder of derivation coincides with that for the sequential update
and we can write for $P(X,t|X^{0},0)$ the determinant formula with the
matrix elements 
\begin{equation}
M_{i,j}=\tilde{F}_{i-j}(x_{i}-x_{j}^{0}|t) \qquad i,j=1,2,...,P  \label{MM1P}
\end{equation}%
The determinant formula for $P(X,t|X^{0},0)$ is valid for all initial states 
$X^{0}$ and for all final states $X$ conditioned by inequalities $%
x_{i+1}-x_{i}\geq 1$ for all $i=1,2,...,P-1$. If $x_{i+1}-x_{i}=1$ for some $%
1\leq i\leq P-1$, the determinant formula should be corrected. To find the
correction, consider the final positions of particles $x_{1}=x,x_{2}=x+1$ at
time $t$ in the two-particle case $P=2$. For the last time step, the final
positions are simultaneously the reference points of unallowed and auxiliary
sets of trajectories. However, these sets cannot be superposed as in the
case of bulk scattering because the range of auxiliary set exceeds the range
of unallowed set (the points $(x,x+2)$ and $(x+1,x+2)$ in Fig.(\ref%
{diagram(abc)}) are absent). Explicit calculations for the transitions from $%
(x-1,t-1),(x+1,t-1)$ to $(x,t),(x+1,t)$ and from $(x,t-1),(x+1,t-1)$ to $%
(x,t),(x+1,t)$ show that the correct probability differs from the
determinant result by the factor $(1-v)$. Using the fact that the scattering
of each pair of trajectories can be considered independently, we write the
final result as 
\begin{equation}
P(X,t|X^{0},0)=(1-v)^{n}\det\mathbf{M},  \label{detfin}
\end{equation}%
where $n$ is the number of pairs of neighboring particles in the final
position. The equivalence of the results (\ref{P(X,t|X_0,0)}-\ref{f(a,b,t)})
and (\ref{detfin}) is verified directly by expanding the hypergeometric
functions in (\ref{f(a,b,t)}).

\begin{figure}[t]
\unitlength=1mm \makebox(120,75)[cc] {\psfig{file=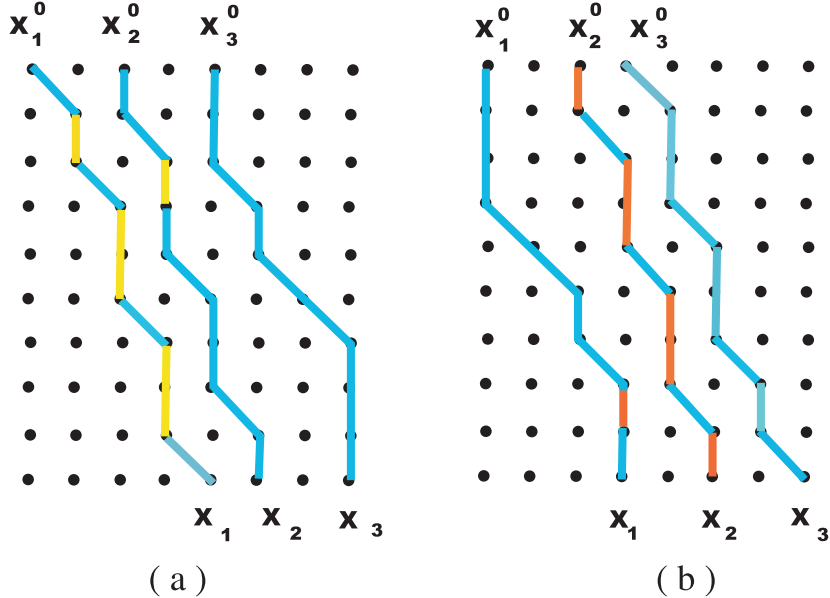,width=140mm}}
\caption{{(a) The initial state contains no pairs of neighboring sites; the
final state contains one pair. (b) The reversed dynamics. The yellow and red
bonds have weight 1. The difference between their numbers equals to the
number of pairs in the final state in (a)}}
\label{infin}
\end{figure}

To make above arguments more transparent, consider a transformation $%
X\Leftrightarrow X^{0}$ of a realization of the process interchanging the
initial and final configurations of particles. As a result, we obtain the
same realization overturned (Fig.\ref{infin}a,b). 

The initial positions in Fig.(\ref{infin}a) and final positions in Fig.(\ref%
{infin}b) obey the conditions $x_{i+1}-x_{i}\geq 1$ for all $i=1,2,...,P-1$.
The final positions in Fig.(\ref{infin}a) and initial positions in Fig.(\ref%
{infin}b) contain pairs of neighboring coordinates. A remarkable property of
the path configurations is that the difference in numbers of bonds having
weight 1 (yellow bonds in Fig.(\ref{infin}a) and red bonds in Fig.(\ref%
{infin}b)) equals to the difference in numbers of pairs in final and initial
positions. Therefore, we can apply the pure determinant formula to the case
when the final position is free of neighboring pairs and then calculate the
opposite case multiplying the determinant by $(1-v)^{n}$ as in (\ref{detfin}%
).

\section{\protect\bigskip Discussion and conclusion.}

To summarize, we considered the discrete time asymmetric exclusion process
with parallel update at the infinite 1-D lattice. We have presented two
alternative derivations for the probability of a particle configuration at
arbitrary time, given an initial configuration. In the first approach we
constructed the transformation of the evolution operator, which allows one
to find its left and right eigenvectors in the Bethe ansatz form. We proved
the completeness of the eigenbasis, which allows calculation of the matrix
elements of an arbitrary power of evolution operator. The second approach is
based on the consideration of the ensembles of trajectories of
noninteracting particles. Taken with alternating signs their statistical
weights are chosen in such a way that they add up to reproduce correctly the
dynamics of interacting particles.

The result for the system of $P$ particles is expressed in the form of the
determinant of $P\times P$ matrix, whose elements are given in terms of the
Gauss hypergeometric functions $_{2}F_{1}(a,b,c;x)$. To relate this result
to the previous ones obtained for the ASEP, we note that in the continuous
time limit $t\rightarrow \infty ,$ $\lambda \rightarrow 0,$ $t\lambda =\tau $
the hypergeometric function $_{2}F_{1}(a,b,c;x)$ degenerates into the
confluent hypergeometric function yielding 
\begin{equation}
f(a,b,\tau )=\left( -1\right) ^{a}e^{-\tau } 
\begin{cases}
\begin{array}{cc}
\frac{\tau ^{b}}{b!}\left. _{1}F_{1}\right. \left( a,b+1,\tau \right) & b>0
\\ 
\frac{\left( a\right) _{-b}}{\left( -b\right) !}\left. _{1}F_{1}\right.
\left( a-b,-b+1,\tau \right) & b\leq 0%
\end{array}%
\end{cases}
,
\end{equation}%
which is equivalent to the results obtained by Sch\"{u}tz, \cite{schutz}.
The relation of Sch\"{u}tz's functions to the confluent hypergeometric
functions was already noticed before \cite{sas nag}.

We should also remark that there is a direct relation of our results to the
determinant solution for the TASEP with backward ordered update \cite%
{brankov priezzhev shelest} referred also to as a fragmentation model \cite%
{rakos schutz}. The determinant formula for that process can be obtained
from our one by using $(1-v)^af(a,b,t-a)$ instead of $f(a,b,t)$ in the
formula for the matrix elements (\ref{F(x,y,t)_i.j}). This connection
between conditional probabilities seems to be similar to the correspondence
noted in \cite{rakos schutz} on the level of the current distributions.
Given such a close relation between the solutions of the master equations,
an establishing of the direct mapping between the two original processes
would be interesting.

The obtained results open rich perspectives for derivation of more
complicated correlation functions for the density and current of particles.
Many of such quantities have been obtained recently due to close connection
of the subject with the statistics of random permutations and ensembles of
random matrices. Particularly, using the mapping of the ASEP with parallel
update to the growing Young diagrams Johansson obtained the probability
distribution of the integrated particle current through an arbitrary bond
starting from the half filled lattice \cite{Johansson}. Properly rescaled
this distribution converges to the distribution of the largest eigenvalue of
the random matrix of the Gaussian unitary ensemble \cite{mehta}. The
limiting case of the Johansson's formula for the continuous time ASEP as
well as its analogue for the ASEP with the backward update were reproduced
using the Bethe ansatz in \cite{rakos schutz}. A derivation of the
Johansson's formula starting from our determinant expression can be proposed
as a closest goal for further investigation. These results were generalized
for other initial conditions, which turned out to be related with different
ensembles of random matrices \cite{Praehofer Spohn}, \cite{sas nag}. In
addition more intricate quantities were considered like multipoint joint
distribution of the integrated current \cite{Sasamoto}, \cite{ferrari spohn1}
or correlation functions of site occupation numbers \cite{ferrari spohn2}.
Most of results were obtained either for continuous time ASEP or for the
ASEP with the backward update. One can hope that the results of present
paper can serve as a starting point for the further studies of the case of
parallel update.

Another direction of research is the generalization of the results for the
ring geometry. In the present article, the proof of the completeness of the
Bethe ansatz basis for evolution operator of the parallel ASEP has been
given for the case of the infinite chain. Similar proof of the completeness
and orthonormality for the case of the ring would be an important issue not
only for the stochastic processes community but in the general context of
the Bethe ansatz solvable models \cite{dorlas}.

\acknowledgements We are grateful to G. Sch\"utz for useful discussions.
This work was supported by the RFBR grant 06-01-00191a

\appendix

\section{Proof of the Proposition \protect\ref{prop}.}

The integral (\ref{integral1}) can be evaluated using the Cauchy theorem. To
this end, one can expand the expression in the parentheses into the Laurent
series. Practically one can first expand $\left( 1+\lambda z\right)
^{i-\sigma _{i}}$ and $\left( 1-1/z\right) ^{\sigma _{i}-i}$ separately and
then evaluate their product. Then the coefficient of the term of power $%
\left( x_{i}-y_{\sigma _{i}}\right) $ will be the result. The choice of the
integration contour means, that the expansion of $\left( 1+\lambda z\right)
^{i-\sigma _{i}}$ will contain only nonnegative powers of $\lambda z$ and
the expansion of $\left( 1-1/z\right) ^{\sigma _{i}-i}$ will contain only
nonnegative powers of $1/z$. In addition we should take into account that
the number of terms of the expansion will be finite when the exponent is
positive. Thus, when 
\begin{equation}
i=\sigma _{i},
\end{equation}
the integral to be nonzero we should require%
\begin{equation}
x_{i}=y_{i}.
\end{equation}%
In the other cases, the following inequalities should hold. 
\begin{eqnarray}
i-\sigma _{i} &\geq &x_{i}-y_{\sigma _{i}},\qquad \mathrm{for\quad }i>\sigma
_{i}  \label{2.2a} \\
\sigma _{i}-i &\geq &y_{\sigma _{i}}-x_{i},\qquad \mathrm{for\quad }i<\sigma
_{i}.  \label{2.2b}
\end{eqnarray}%
Moreover the domain (\ref{domain}), where the coordinates of particles are
defined, implies 
\begin{eqnarray}
i-\sigma _{i} &\leq &x_{i}-x_{\sigma _{i}},\mathrm{\quad }i-\sigma _{i}\leq
y_{i}-y_{\sigma _{i}},\qquad \mathrm{for\quad }i>\sigma _{i}  \label{2.5a} \\
\sigma _{i}-i &\leq &x_{\sigma _{i}}-x_{i},\mathrm{\quad }\sigma _{i}-i\leq
y_{\sigma _{i}}-y_{i},\qquad \mathrm{for\quad }i<\sigma _{i}.  \label{2.5b}
\end{eqnarray}%
Comparing the expressions (\ref{2.2a},\ref{2.2b}) with (\ref{2.5a},\ref{2.5b}%
) respectively, we obtain%
\begin{eqnarray}
y_{\sigma _{i}} &\geq &x_{\sigma _{i}},\mathrm{\quad }y_{i}\geq x_{i}\qquad 
\mathrm{for\quad }i>\sigma _{i}  \label{2.6a} \\
y_{\sigma _{i}} &\leq &x_{\sigma _{i}},\mathrm{\quad }y_{i}\leq x_{i}\qquad 
\mathrm{for\quad }i<\sigma _{i}.  \label{2.6b}
\end{eqnarray}%
\qquad

Any permutation can be decomposed into the set of disjoint orbits \cite%
{comtet}, i.e. the sequences of elements which trade places with one another
under a given permutation. Consider an orbit $i_{1}\rightarrow
i_{2}\rightarrow \cdots \rightarrow i_{k}\rightarrow i_{1}$, such that 
\begin{equation}
\sigma _{i_{1}}=i_{2},\sigma _{i_{2}}=i_{3},\cdots ,\sigma _{i_{k}}=i_{1}
\label{2.7}
\end{equation}%
There exists $r$ such that $i_{r}$ is the largest in a given orbit. Then
from $i_{r}>i_{r+1}\equiv \sigma _{i_{r}}$ and the second inequality of (\ref%
{2.6a}) we have%
\begin{equation}
y_{i_{r}}\geq x_{i_{r}},
\end{equation}%
while from $i_{r-1}<i_{r}\equiv \sigma _{i_{r-1}}$ and the first inequality
of (\ref{2.6b}) it follows that 
\begin{equation}
y_{i_{r}}\leq x_{i_{r}}.
\end{equation}%
These two are consistent only if 
\begin{equation}
y_{i_{r}}=x_{i_{r}}.  \label{2.12}
\end{equation}%
The latter equality together with the inequality (\ref{2.2a}) and the second
inequality of (\ref{2.5a}) yield 
\begin{equation}
y_{i_{r}}-y_{i_{r+1}}\geq i_{r}-i_{r+1}\geq y_{i_{r}}-y_{i_{r+1}},
\end{equation}%
which means 
\begin{equation}
i_{r}-i_{r+1}=y_{i_{r}}-y_{i_{r+1}}.
\end{equation}%
In the same way using the inequalities (\ref{2.2a},\ref{2.5a}) we obtain%
\begin{equation}
i_{r}-i_{r-1}=x_{i_{r}}-x_{i_{r+1}}.
\end{equation}%
We should note that the equality like 
\begin{equation}
x_{i}-x_{j}=i-j
\end{equation}%
means that there are as many sites between the sites $x_{i}$ and $x_{j}$ as
many particles. In other words $i$-th and $j$-th particles belong to the
same cluster. Apparently, the coordinates of any pair of particles between
them also satisfy similar relation. One can see that the particles at sites $%
y_{i_{r}},y_{i_{r+1}}$ belong to the same cluster of the configuration $Y$,
while the particles at sites $x_{i_{r}},x_{i_{r-1}}$ belong to the same
cluster of the configuration $X$. Let $i_{r+1}$ be the largest of the
numbers $i_{r+1},$ $i_{r-1}$. The cluster of particles spreads from $%
x_{i_{r}}$ backward to at least $x_{i_{r+1}}$ in both configurations $X$ and 
$Y$. Therefore we conclude that 
\begin{equation}
x_{i_{r+1}}=y_{i_{r+1}}.
\end{equation}%
Then we look for $i_{r+2}$. If $i_{r+2}$ is larger than $i_{r+1}$, then it
is between $i_{r}$ and $i_{r+1}$ and hence belongs to the same cluster in
both $X$ and $Y$ and we should look for $i_{r+3}$. If $i_{r+2}$ is less than 
$i_{r+1}$, using the same arguments as for $i_{r}$, $i_{r+1}$ we have 
\begin{equation}
y_{i_{r+1}}-y_{i_{r+2}}=i_{r+2}-i_{r+1},
\end{equation}%
i.e. we extend the cluster in $Y$ backward up to \thinspace $i_{r+2}$-th
particle. \ Then we again compare \thinspace $i_{r+2}$ with $i_{r-1}$, which
currently limits the cluster in $X$ from behind. If $i_{r-1}$ is between $%
i_{r+1}$ and $i_{r+2}$ it necessarily belongs to the same cluster also in $Y$%
. Then the cluster can be extended up to $i_{r-2}$, etc. Thus, step by step
we extend the cluster backward in $Y\,$going forward along the orbit and
backward in $X$ going backward along the orbit until we come to the same
element in both directions, which means that the orbit is exhausted. Finally
we conclude that all the particles in a given orbit should belong to the
same cluster and their coordinates in the configurations $X$ and $Y$
coincide.

\section{Explicit form of auxiliary sets}

The transformation $\mathbb{\ V} \Rightarrow \mathbb{\ A}_{v}(k_1,k_2)$ is

$[(x_{1}^{0},0)\rightarrow (x_{c},t_{c})\rightarrow
(x_{c}+1,t_{c}+1)\rightarrow (x_{1},t)]\otimes \lbrack
(x_{2}^{0},0)\rightarrow (x_{c}+1,t_{c})\rightarrow
(x_{c}+1,t_{c}+1)\rightarrow (x_{2},t)]\Rightarrow \lbrack
(x_{1}^{0}-k_{1},0)\rightarrow (x_{c}-k_{1},t_{c})\rightarrow
(x_{c}+1-k_{1},t_{c}+1)\rightarrow (x_{2},t)]\otimes \lbrack
(x_{2}^{0}-k_{2},0)\rightarrow (x_{c}+1-k_{2},t_{c})\rightarrow
(x_{c}+1-k_{2},t_{c}+1)\rightarrow (x_{1},t)]$,

and the transformation $\mathbb{\ W}\Rightarrow \mathbb{\ A}_{w}(k_{1},k_{2})
$ is

$[(x_1^0,0)\rightarrow (x_c,t_c) \rightarrow (x_c,t_c+1) \rightarrow
(x_1,t)] \otimes [(x_2^0,0)\rightarrow (x_c+1,t_c) \rightarrow (x_c+1,t_c+1)
\rightarrow (x_2,t)] \Rightarrow [(x_1^0-k_1,0)\rightarrow (x_c-k_1,t_c)
\rightarrow (x_c-k_1,t_c+1) \rightarrow (x_2,t)] \otimes
[(x_2^0-k_2,0)\rightarrow (x_c+1-k_2,t_c) \rightarrow (x_c+1-k_2,t_c+1)
\rightarrow (x_1,t)]$.

It follows from the definitions that 
\begin{equation}
\bigcup_{(x_c,t_c)\in \Lambda}\bigcup_{k_1=0}^{\infty}\bigcup_{k_2=0}^{1}
\left( \mathbb{\ A}_{v}(k_1,k_2)\cup\mathbb{\ A}_{w}(k_1,k_2) \right) = 
\mathbb{\ A}_{21}.  \label{normA21}
\end{equation}

The weights of introduced sets obey the following relations 
\begin{eqnarray}
\mu \left( \mathbb{\ A}_{v}(0,0)\right) &=&\mu \left( \mathbb{\ V}%
(x_{c},t_{c})\right) , \\
\mu \left( \mathbb{\ A}_{v}(0,1)\right) &=&-\mu \left( \mathbb{\ W}%
(x_{c},t_{c})\right) , \\
\mu \left( \mathbb{\ A}_{w}(k,0)\right) &=&-\mu \left( \mathbb{\ A}%
_{v}(k+1,0)\right) ,\;k=0,1,2\ldots \;, \\
\mu \left( \mathbb{\ A}_{w}(k,1)\right) &=&-\mu \left( \mathbb{\ A}%
_{v}(k+1,1)\right) ,\;k=0,1,2\ldots \;.
\end{eqnarray}

Transformation $\mathbb{\ X} \Rightarrow \mathbb{\ A}_{x}(k_1,k_2)$ is

$[(x_{1}^{0},0)\rightarrow (x_{c},t_{c})\rightarrow
(x_{c}+1,t_{c}+1)\rightarrow (x_{1},t)]\otimes \lbrack
(x_{2}^{0},0)\rightarrow (x_{c}+1,t_{c})\rightarrow
(x_{c}+2,t_{c}+1)\rightarrow (x_{2},t)]\Rightarrow \lbrack
(x_{1}^{0}-k_{1},0)\rightarrow (x_{c}-k_{1},t_{c})\rightarrow
(x_{c}+1-k_{1},t_{c}+1)\rightarrow (x_{2},t)]\otimes \lbrack
(x_{2}^{0}-k_{2},0)\rightarrow (x_{c}+1-k_{2},t_{c})\rightarrow
(x_{c}+2-k_{2},t_{c}+1)\rightarrow (x_{1},t)]$\newline

and the transformation $\mathbb{\ Y}\Rightarrow \mathbb{\ A}_{y}(k_{1},k_{2})
$ is

$[(x_1^0,0)\rightarrow (x_c,t_c) \rightarrow (x_c,t_c+1) \rightarrow
(x_1,t)] \otimes [(x_2^0,0)\rightarrow (x_c+1,t_c) \rightarrow (x_c+2,t_c+1)
\rightarrow (x_2,t)] \Rightarrow [(x_1^0-k_1,0)\rightarrow (x_c-k_1,t_c)
\rightarrow (x_c-k_1,t_c+1) \rightarrow (x_2,t)] \otimes
[(x_2^0-k_2,0)\rightarrow (x_c+1-k_2,t_c) \rightarrow (x_c+2-k_2,t_c+1)
\rightarrow (x_1,t)]$.

\end{document}